\documentclass[twocolumn]{IEEEtran}
\usepackage{amsmath,amssymb,amsthm,epsfig,color,subfigure,empheq,graphicx,graphics,balance}
\usepackage{enumerate,url,algorithm,algorithmic,wasysym,epstopdf,enumitem}
\usepackage{xcolor}
\usepackage{amsfonts}
\usepackage{adjustbox}
\usepackage{multirow}

\usepackage{makecell}
\usepackage{pifont}
\newcommand{\xmark}{\ding{55}}%

\DeclareMathOperator{\diag}{dg}

\DeclareMathOperator{\vectorize}{vec}

\newtheorem{assumption}{Assumption}
\newtheorem{proposition}{Proposition}

\newtheorem{remark}{Remark}

\newcommand \bzero{\mathbf{0}}
\newcommand \bone{\mathbf{1}}

\newcommand \bc{\mathbf{c}}

\newcommand \be{\mathbf{e}}

\newcommand \bp{\mathbf{p}}

\newcommand \bx{\mathbf{x}}
\newcommand \by{\mathbf{y}}
\newcommand \bz{\mathbf{z}}
\newcommand \bA{\mathbf{A}}
\newcommand \bB{\mathbf{B}}
\newcommand \bC{\mathbf{C}}
\newcommand \bD{\mathbf{D}}

\newcommand \bH{\mathbf{H}}
\newcommand \bI{\mathbf{I}}

\newcommand \bK{\mathbf{K}}
\newcommand \bL{\mathbf{L}}
\newcommand \bM{\mathbf{M}}

\newcommand \bP{\mathbf{P}}

\newcommand \bR{\mathbf{R}}
\newcommand \bS{\mathbf{S}}
\newcommand \bT{\mathbf{T}}
\newcommand \bU{\mathbf{U}}
\newcommand \bV{\mathbf{V}}

\newcommand \bX{\mathbf{X}}
\newcommand \bY{\mathbf{Y}}

\newcommand \balpha{\boldsymbol{\alpha}}

\newcommand \btheta{\boldsymbol{\theta}}

\newcommand \bmu{\boldsymbol{\mu}}

\newcommand \bsigma{\boldsymbol{\sigma}}

\newcommand \bomega{\boldsymbol{\omega}}

\newcommand \bTheta{\mathbf{\Theta}}

\newcommand \bLambda{\mathbf{\Lambda}}

\newcommand \bSigma{\mathbf{\Sigma}}

\newcommand \bOmega{\mathbf{\Omega}}

\newcommand \mcM{\mathcal{M}}
\newcommand \mcN{\mathcal{N}}

\newcommand \mcT{\mathcal{T}}

\newcommand \tbk{\tilde{\mathbf{k}}}

\newcommand \tby{\tilde{\mathbf{y}}}

\newcommand \tbA{\tilde{\mathbf{A}}}

\newcommand \tbK{\tilde{\mathbf{K}}}

\newcommand \tbalpha{\tilde{\boldsymbol{\alpha}}}

\begin{document}

\title{Inferring Power System Dynamics from Synchrophasor Data using Gaussian Processes}

\author{
	Mana Jalali,~\IEEEmembership{Student Member,~IEEE},
	Vassilis Kekatos,~\IEEEmembership{Senior Member,~IEEE}, 
	Siddharth Bhela,~\IEEEmembership{Member,~IEEE},
	Hao Zhu,~\IEEEmembership{Senior Member,~IEEE}, and	
	Virgilio Centeno,~\IEEEmembership{Senior Member,~IEEE}


\thanks{This work was supported by the U.S. National Science Foundation through grants 1751085 and 1923221. M.~Jalali, V.~Kekatos, and V.~Centeno are with the Bradley Dept. of ECE, Virginia Tech, Blacksburg, VA 24061, USA. S.~Bhela is with Siemens Corporate Technology, Princeton, NJ 08540, USA. H.~Zhu is with the Dept. of ECE, the University of Texas at Austin, Austin, TX 78712, USA. Emails:\{manaj2,kekatos,virgilio\}@vt.edu, siddharth.bhela@siemens.com, haozhu@utexas.edu}	
}	
	

\maketitle

\begin{abstract}
Synchrophasor data provide unprecedented opportunities for inferring power system dynamics, such as estimating voltage angles, frequencies, and accelerations along with power injection at all buses. Aligned to this goal, this work puts forth a novel framework for learning dynamics after small-signal disturbances by leveraging Gaussian processes (GPs). We extend results on learning of a linear time-invariant system using GPs to the multi-input multi-output setup. This is accomplished by decomposing power system dynamics into a set of single-input single-output linear systems with narrow frequency pass bands. The proposed learning technique captures time derivatives in continuous time, accommodates data streams sampled at different rates, and can cope with missing data and heterogeneous levels of accuracy. While Kalman filter-based approaches require knowing all system inputs, the proposed framework handles readings of system inputs, outputs, their derivatives, and combinations thereof collected from an arbitrary subset of buses. Relying on minimal system information, it further provides uncertainty quantification in addition to point estimates of system dynamics. Numerical tests verify that this technique can infer dynamics at non-metered buses, impute and predict synchrophasors, and locate faults under linear and non-linear system models under ambient and fault disturbances.
\end{abstract}
	

\allowdisplaybreaks

\section{Introduction}\label{sec:intro}
Maintaining the stability and synchronization of a power system can be enhanced upon closely monitoring the voltage angles, frequencies, accelerations (rates of change of frequency or ROCOF), as well as the power injections at all buses. Phasor measurement units (PMUs) provide high-accuracy data on dynamic system states at high temporal resolution. However, due to high installation and networking costs, not all buses are instrumented with PMUs, while communication failures oftentimes result in missing PMU readings~\cite{PMUcost}. Such challenges motivate the need for inferring power system dynamics from synchrophasor data using minimal system information. Since observability may not be always granted, measures for quantifying the uncertainty are relevant. We propose a framework to estimate the signals involved in power system dynamics after small-signal disturbances. The suggested framework can be employed for various applications such as data imputation and screening, frequency monitoring, and localization of oscillations, to name a few.

Approaches to infer power system dynamics can be broadly classified into data- and model-based methods. Data-based methods typically use synchrophasor measurements to learn the system's dynamic states. Reference~\cite{chow14} for example advocates that the matrix collecting PMU measurements across buses and time instances features a low-rank plus sparse structure, so missing PMU data could be recovered by means of matrix completion~\cite{chow16}. If all PMU data is lost for one or more consecutive time instances, a robust matrix completion approach stacking data in a Hankel matrix shows promise to recover the original PMU data stream~\cite{RMC_Wang19},~\cite{Wang18}; though performance deteriorates with prolonged periods of lost communication. The work in~\cite{Chaud20} proposes grouping the measured signals prior to robust principal component analysis to meet the sufficient conditions of guaranteed data recovery. Arranging synchrophasor readings in higher-order tensors rather than matrices could potentially resolve the latter issue using tensor decomposition techniques~\cite{Chow20_tensor}. Nonetheless, data-based techniques cannot extrapolate on buses not instrumented with PMUs, do not utilize readings of power injections or flows, and ignore any system model information. 


Dynamic state estimation (DSE) aims at inferring the power system states using both a system model and measurements processed through a Kalman filter (KF); see~\cite{DSE19} for a recent comprehensive review. Plain KFs are optimal estimators that adopt a linear system model, while nonlinear power system dynamics can be handled through KF variants, such as the extended
~\cite{Huang02}; the unscented~\cite{ulmann04}, \cite{Meliopoulos12}; and ensemble KFs~\cite{Huang18}, \cite{SLu13}. Despite these developments in sophisticating system models, KF-based DSE solutions operate on a localized fashion and consider a singe bus or a single control area of the power system~\cite{Pal14}, \cite{Mili18}. More importantly, KF-based methods presume all inputs to the dynamic system of interest (single bus or control area of the power system) are either known or measured. Such requirement may not be realistic for the entire power system. Moreover, typically KFs operate on data collected over uniformly sampled time intervals, which renders them vulnerable to missing data or different sampling rates. Finally, DSE approaches approximate continuous differential equations with discrete finite differences. 

Synchrophasor data can be used to infer more coarse dynamic system information, such as locating the sources of oscillations. The latter task can be accomplished by comparing the arrival time of traveling waves \cite{thorp98} and \cite{YLiu-map}; by measuring the dissipated energy of power flows~\cite{Liu_ISGT19}, \cite{Litvinov}; or via robust principal component analysis~\cite{le_xie20}. But these methods require access to data across the entire power network. 

In a nutshell, existing methods for inferring power system dynamics have limitations on the placement and sampling rate of measurements, while data streams should be reliable and uninterrupted. To overcome these restrictions, we propose a comprehensive framework for learning power system dynamics from PMU data using Gaussian processes (GPs). Our methods rely on approximate system information, such as the inertia parameters for generators and the Jacobian matrix of the power flow equations. A detailed comparison with existing works is deferred to Section~\ref{subsec:comparison}.

Our contribution is threefold: \emph{i)} Cross-pollinate results for GP-based inference on a single-input single-output (SISO) linear time-invariant (LTI) system to the task of learning power system dynamics; \emph{ii)} Leverage the physics behind the swing equations and inter-area oscillations to extend GP modeling from the SISO to the multi-input multi-output (MIMO) setup and infer dynamics at non-metered buses; and \emph{iii)} Develop a scalable technique for estimating GP model parameters from collected data using the method of moments. 

The proposed toolbox comes with several unique features. First, signals are modeled in a continuous fashion, which lends itself a natural way to compute time derivatives, which is robust even under low sampling rates and missing data. Second, system inputs and outputs are handled in a unified manner: Power injections, voltage angles, frequencies, and ROCOFs at any bus can be treated either as measured or wanted signals without major changes to the framework. Finally, thanks to its Bayesian flavor, the GP inference paradigm provides not only a point estimate, but also a Gaussian probability distribution function (PDF) for the sought signal. The latter feature allows for uncertainty quantification for the estimated data streams. This is important when testing data normality against attacks or under limited observability.

The rest of the paper is organized as follows. Section~\ref{sec:problem} defines the general problem setup of \emph{learning power system dynamics}. Section~\ref{sec:GP} reviews Gaussian processes and adopts them to learning in linear dynamical systems. Section~\ref{sec:power} builds on the swing equation to develop a statistical model for power system dynamics. Section~\ref{sec:inference} proposes a model reduction method for increasing the efficiency of the GP paradigm and estimating the needed parameters. It also contrasts our new GP-based learning methodology to existing works. Our methodology is tested under various application setups in Section~\ref{sec:tests}. Conclusions and future directions are outlined in Section~\ref{sec:conc}.

\emph{Notation:} column vectors (matrices) are denoted by lower- (upper-) case letters. Operator $\diag(\bx)$ returns a diagonal matrix with $\bx$ on its main diagonal. Symbol $(\cdot)^{\top}$ stands for transposition; $\bI_{N}$ is the $N \times N$ identity matrix; $\dot{x}=\frac{dx}{dt}$ denotes time differentiation; and $\mathbb{E}$ is the expectation operator. Operator $\mathrm{vec}(\bX)$ vectorizes a matrix by stacking its columns in a single vector, while $[\bX]_{i,j}$ is the $(i,j)$-th entry of $\bX$. The Kronecker product of matrices is expressed as $\bX \otimes \bY$. The notation $\bx\sim \mcN(\bmu,\bSigma)$ means $\bx$ follows a multivariate Gaussian distribution with mean $\bmu$ and covariance $\bSigma$.

\section{Problem Statement and Relevant Applications}\label{sec:problem}
We are interested in monitoring power system dynamics under small-signal disturbances using synchrophasor data. Power system dynamics are modeled here through an approximate multi-input multi-output (MIMO) linear time-invariant (LTI) system. The \emph{inputs} to this MIMO LTI system are the deviations from the scheduled active power injections. Its \emph{outputs} or \emph{states} correspond to deviations from the steady-state rotor angles or speeds, denoted by $\theta_n(t)$ and $\omega_n(t)=\dot{\theta}_n(t)$ per bus $n$. For brevity, we henceforth drop the term \emph{deviations}. The rate-of-change-of-frequency (ROCOF) or acceleration $\dot{\omega}_n(t)$ may be measured or may be of interest. The angle, speed, and acceleration of a synchronous machine is captured by the angle, speed (frequency), and acceleration of the related voltage phasor. To avoid confusion between frequency $\omega_n(t)$ and the frequency-domain analysis of time signals, we will henceforth refer to $\omega_n(t)$ as \emph{speed}. The parameters of the aforesaid MIMO LTI system are assumed to be known, precisely or approximately~\cite{dosiek13}, \cite{Zhu18}. These parameters include generator constants (e.g., inertia and damping) as well as the Jacobian matrix of the power flow equations evaluated at the current operating point or the flat voltage profile.

The envisioned application setup is described next.  The system operator is collecting synchronized data of voltage angles, speeds, ROCOFs, or power injections on a subset of buses. The collected data may be of different degrees of accuracy due to instrumentation or estimation noise.
The goal is to infer non-metered grid quantities related to power system dynamics. The collected data are noisy and sampled partially across buses and time. They are also heterogeneous since they may include system inputs, outputs, and their derivatives. 

The proposed learning framework can be used in different application setups, such as: \emph{i)} Given voltage angles at some buses, monitor the speeds at non-metered buses to ensure stability; \emph{ii)} Given voltage angles, speeds, and power injections at generator buses (all or a subset of them), infer the power injections at the remaining buses to localize faults or sources of oscillations; \emph{iii)} Impute missing entries from a synchrophasor data stream or cross-validate a data stream that has been deemed erroneous or suspicious; and \emph{iv)} Compute reliable estimates for speeds and ROCOFs to drive load-frequency control and grid-forming inverters. 


\section{Gaussian Processes for Learning in Dynamical Systems}\label{sec:GP}
This section reviews the basics of Gaussian processes and explains how GPs can model single-input single-output (SISO) LTI dynamical systems. A GP is a random process with the additional property that any collection of a finite number of its samples forms a Gaussian random vector~\cite[Ch.1]{GP}. Consider for example a time series $x(t)$ and two distinct sets of time indices $\mcT_1$ and $\mcT_2$. A signal $x(t)$ is a GP if the two vectors $\bx_1$ and $\bx_2$ collecting samples of $x(t)$ over $\mcT_1$ and $\mcT_2$ are jointly Gaussian or
\begin{equation}\label{eq:jointGP}
\begin{bmatrix}
\bx_1\\
\bx_2
\end{bmatrix} \sim \mcN \left( \begin{bmatrix}
\bmu_1\\
\bmu_2
\end{bmatrix} , \begin{bmatrix}
\bSigma_{11} & \bSigma_{21}^\top\\
\bSigma_{21} & \bSigma_{22}
\end{bmatrix}\right).
\end{equation}
Due to~\eqref{eq:jointGP}, the conditional probability density function (PDF) of $\bx_2$ given $\bx_1$ is also Gaussian with mean and covariance~\cite[Ch.~6.4]{Bishop}
\begin{subequations}\label{eq:x2}
	\begin{align}
	\mathbb{E}[\bx_2|\bx_1]&=\bmu_2 + \bSigma_{21}\bSigma_{11}^{-1} (\bx_1 -\bmu_1)\label{eq:x2:m}\\
	\mathrm{Cov}[\bx_2|\bx_1]&=\bSigma_{22} - \bSigma_{21} \bSigma_{11}^{-1} \bSigma_{21}^\top.\label{eq:x2:c}
	\end{align}
\end{subequations}
Such modeling is useful because knowing the value of $\bx_1$, the minimum mean square error (MMSE) estimator of $\bx_2$ is~\eqref{eq:x2:m}. Moreover, the uncertainty of this estimate is described by~\eqref{eq:x2:c}. 

In the learning problems to be addressed in this work, we do not indent to make any observability claim and determine whether the dynamic signals of interest in $\bx_2$ can be observed from the given measurements in $\bx_1$. Nonetheless, thanks to the Bayesian nature of the approach, the diagonal entries of $\mathrm{Cov}[\bx_2|\bx_1]$ can be used as confidence intervals. As an example, under limited number of measurements or when $\bx_2$ cannot be observed from $\bx_1$, the covariance $\mathrm{Cov}[\bx_2|\bx_1]$ would take large values on its diagonal entries, which prompt us that $\mathbb{E}[\bx_2|\bx_1]$ is not a reliable estimate of $\bx_2$. The statistical characterization of $\bx_2$ in \eqref{eq:x2} can also be utilized for anomaly detection. Suppose that $\bx_2$ is actually observed, but is not trustworthy. One can use a trustworthy $\bx_1$ and the formulae in \eqref{eq:x2} to compute the Gaussian conditional PDF $p(\bx_2|\bx_1)$. If this PDF takes on a relatively small value when evaluated at the actual $\bx_2$, then $\bx_2$ can be flagged as anomalous.

Let us review how GPs can be used for learning in LTI systems; see~\cite{raissi17},~\cite{Graepel} for details. Consider the SISO LTI system described by the ordinary differential equation (ODE)
\begin{equation}\label{eq:siso}
\ddot{y}(t) + \gamma \dot{y}(t) + \lambda y(t) = x(t)
\end{equation}
for given $\gamma,\lambda>0$, and initial conditions $y(0)$ and $\dot{y}(0)$. If $x(t)$ is known, then $y(t)$ can be computed as the solution of \eqref{eq:siso} using standard ODE methods. For the inverse problem, if the output $y(t)$ is known, its derivatives can be computed and the input $x(t)$ can be found by simple substitution in \eqref{eq:siso}. 
	
Both problems get more complicated when the known signal (input or output) is observed through noisy discrete-time samples. Moreover, for the inverse problem, one may not have access to all derivatives $\{y,\dot{y},\ddot{y}\}$. And perhaps the measurements of $\{x,y,\dot{y},\ddot{y}\}$ are originating from sensors or estimation methods with different levels of accuracy. The goal is to estimate the non-metered signals. Modeling $y(t)$ as a GP provides a statistical framework to do so as explained next.

Let us model $y(t)$ as a zero-mean GP with the covariance $\mathbb{E}[y(t)y(t')]=k(t,t')$. Function $k(t,t')$ is known as the \emph{kernel function} and can be decided based on prior knowledge about the signal. We are interested in the joint PDF for samples of $y(t)$ collected at times $\mcT := \{t_1,\dots,t_T\}$, and stacked in vector $\by:=[y(t_1)~\cdots~y(t_T)]^\top$. Without loss of generality, the sampled times have been ordered as $t_1<\ldots<t_T$. It is not hard to verify that $\by\sim \mcN(\bzero,\bK)$ where $\bK \succeq 0$ and $[\bK]_{ij}=k(t_i,t_j)$ for all $t_i,t_j \in \mcT$. Since the latter holds for any collection of time instances $\mcT$, signal $y(t)$ is a GP indeed. 

Let us focus on the covariance matrix $\bK$ of $\by$. Despite $\bK$ being dependent on $\mcT$, our notation drops that dependence for simplicity. The choice of $k(t_i,t_j)$ is crucial for modeling $y(t)$. A signal that is a linear function of $t$ as $y(t)=w_1t$ with $w_1\sim \mcN(0,1)$ possesses the kernel function $k(t_i,t_j)=\mathbb{E}[y(t_i)y(t_j)]=t_it_j$. If a signal is a quadratic function of $t$ as $y(t)=w_2t^2+w_1y+w_0$ with weights being independent zero-mean Gaussian random variables with variances $\mathbb{E}[w_2^2]=\mathbb{E}[w_0^2]=1$ and $\mathbb{E}[w_1^2]=2$, the signal gets the kernel function $k(t_i,t_j)=(1+t_it_j)^2$. The previous two examples explain how the kernel function and hence $\bK$ as $[\bK]_{i,j}=k(t_i,t_j)$, specify the shape of $y(t)$~\cite{Bishop}. A typical choice for kernel function is the Gaussian bell $k(t_i,t_j)=e^{-\beta(t_i-t_j)^2}$ for $\beta >0$, which is appropriate for modeling smooth functions $y(t)$~\cite[Ch.~4]{GP}. Note that GPs have been used in power systems before, e.g., to detect data attacks to steady-state state estimation~\cite{Cyber_GP}, or to predict prices in electricity markets~\cite{KZG14}. Here we leverage the interesting properties of GPs when it comes to time signals to learn grid dynamics.


An appealing property of GPs is that time integration and differentiation of a GP yields a GP~\cite{Graepel}. If the $y(t)$ of \eqref{eq:siso} is a zero-mean GP, then $\dot{y}(t)$ is also a zero-mean GP. If vector $\dot{\by}$ collects the samples of $\dot{y}(t)$ over $\mcT$, define its covariance matrix as
\begin{equation}\label{eq:kdot}
\dot{\bK}:=\mathbb{E} [\dot{\by} \dot{\by}^\top].
\end{equation}
Interestingly, matrix $\dot{\bK}$ can be obtained from $\bK$ as follows:
\begin{equation*}
[\dot{\bK}]_{i,j}=\mathbb{E} \left[\frac{\partial y(t_i)}{\partial t_i}   \frac{\partial y(t_j)}{\partial t_j}\right]=\frac{\partial^2 \mathbb{E}[y(t_i)y(t_j)]}{\partial t_i \partial t_j}=\frac{\partial^2 k(t_i,t_j)}{\partial t_i \partial t_j}.
\end{equation*}
We can similarly show that the $(i,j)$-th entry of $\mathbb{E} [\by \dot{\by}^\top]$ can be computed as $\partial k(t_i,t_j)/\partial t_j$. In general, the covariance between any pair of the signals $(y,\dot{y},\ddot{y})$ appearing in \eqref{eq:siso} can be derived from the kernel function likewise. 


	
Another property of GPs is that a linear combination of GPs is a GP itself: If the state $y(t)$ of the system in \eqref{eq:siso} and its derivatives are GPs, then its input $x(t)$ is a GP as well. In a nutshell, because the kernel $k(t_i,t_j)$ is known analytically, we can readily compute any covariance among $(x,y,\dot{y},\ddot{y})$. 

Armed with a GP model for~\eqref{eq:siso}, several learning scenarios related to the SISO system of \eqref{eq:siso} can be addressed, such as:
\renewcommand{\labelenumi}{\emph{\roman{enumi})}}
\begin{enumerate}
\item \emph{Filtering:} Given samples of $x(t)$, find the state $y(t)$.
\item \emph{Smoothing/prediction:} Given $\by$, infer $y(t')$ for $t'$ within or outside the observation interval $(t_1,t_T)$.
\item \emph{Inverse task:} Given $\by$, find $x(t)$ within or outside $\mcT$.
\item \emph{Mixed setups:} Given samples of $(x,y,\dot{y},\ddot{y})$ (some or all), find the remaining signals at the same or different times. 
\end{enumerate}
In all scenarios, the observed signals may be corrupted by noise and/or sampled at non-uniform intervals. 

\begin{figure*}[t]
\centering
\includegraphics[scale=0.47]{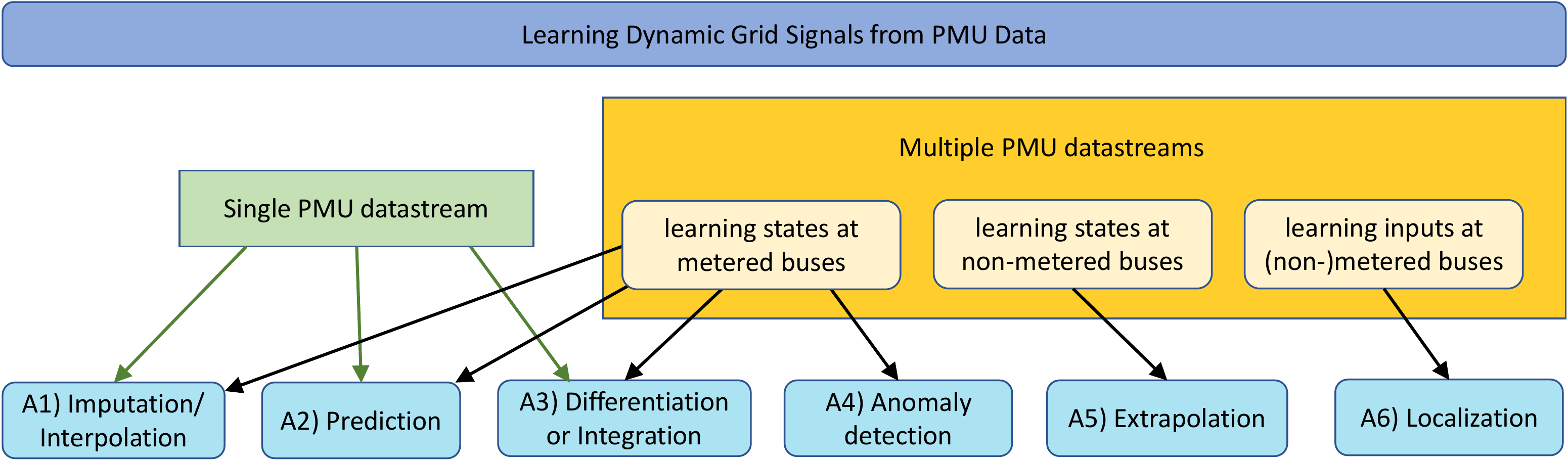}
\caption{Possible applications of processing PMU data for learning dynamic grid signals. \emph{Imputation} refers to finding missing entries, while \emph{interpolation} entails tasks such as upsampling or sampling at different time instances. \emph{Anomaly detection} is possible by having a probabilistic characterization (such as a PDF) for a random signal, so outlying (erroneous or suspicious) instantiations of this signals can be pinpointed. \emph{Extrapolation} is the ability to learn dynamic grid signals on non-metered buses. It can be particularized to any of the applications \emph{A1)}--\emph{A4)}. \emph{Localization} relates to learning the inputs to the dynamic grid model (power injections), and could be useful for unveiling sources of oscillations.}
\label{fig:apps}  
\end{figure*}

The GP toolbox can cope with all tasks in a systematic and unified fashion as long as the covariances appearing in \eqref{eq:x2} are known. We explained earlier that upon postulating a kernel for $y(t)$, all the covariances needed for learning over \eqref{eq:siso} can be readily computed. The previous discussion is unfortunately limited to the SISO case. If we extend this GP-based learning framework to the MIMO setup of power system dynamics, we will be able to handle a gamut of applications, such as those depicted in Fig.~\ref{fig:apps}; note that the applications are described in detail in later sections. To extend GPs from the SISO to the MIMO setup, covariances should be computed across both time and buses. We do so by leveraging swing dynamics as elaborated next.

\section{Modeling Power System Dynamics}\label{sec:power}
The dynamic behavior of a power system can be modeled by a set of nonlinear differential equations in terms of the rotor angles and speeds of the synchronous machines at each bus. Focusing on small-signal analysis, these equations can be linearized around the current operating point yielding the swing equation~\cite{Kundur94}. Consider a power system having $N$ buses hosting synchronous generators comprising set $\mcN$ with rotor angles and speeds collected respectively in $\btheta(t):=[\theta_1(t)~\cdots~\theta_N(t)]^\top$ and $\bomega(t):=[\omega_1(t)~\cdots~\omega_N(t)]^\top$ with $\bomega(t)=\dot{\btheta}(t)$. The mismatch between the electric and mechanical power at each generator is stacked in vector $\bp(t):=[p_1(t)~\cdots~p_N(t)]^\top$.\footnote{In Section~\ref{sec:GP}, vectors collected samples of one signal across time instances $t\in\mcT$. Hereafter, vectors indexed by $t$ collect signals across buses at time $t$.} With these definitions in place, the swing equation can be expressed as~\cite[Ch.~3]{Kundur94}
\begin{equation}\label{eq:swing}
\bM \dot{\bomega}(t) + \bD \bomega(t) + \bL \btheta(t) = \bp(t)
\end{equation}
where $\bM$ and $\bD$ are diagonal matrices collecting the inertia and damping coefficients of generators $M_n$ and $D_n$, and $\bL$ is the negative Jacobian matrix of the power flow equations evaluated at the current operating point and after Kron reduction to remove the effect of non-dynamic buses; see e.g.,~\cite{MKPSCC21},~\cite[Ch.~7]{pai_stability} for details. Within some standard approximations, matrix $\bL$ can be assumed to be symmetric positive semidefinite (psd); see~\cite{Zhu18} for details. 
	
	
As evidenced by \eqref{eq:swing}, grid dynamics can be approximately modeled by a second-order MIMO LTI system. We henceforth select $\bomega$ to be the \emph{state} of this system. It is worth noting that swing dynamics are oftentimes expressed as a first-order dynamical system whose state concatenates $\bomega$ and $\btheta$, and forms its standard state-space representation. Here we intentionally keep swing dynamics in their original second-order form. This will provide intuition, motivate simplifications, and lead to tractable statistical models.

To utilize GP models, we need to be able to compute the covariance $\mathbb{E}[\omega_n(t+\tau)\omega_m(t)]$ for any pair of buses $(n,m)$ and any pair of times $(t+\tau,t)$. There are two challenges here: \emph{i)} The number of such covariance functions grows quadratically with $N$; and \emph{ii)} If bus $n$ is not metered, learning the covariance $\mathbb{E}[\omega_n(t+\tau)\omega_m(t)]$ from data would be impossible even if bus $m$ is metered. To bypass these challenges, the standard approach in GP-based learning is to postulate a parametric form on $\mathbb{E}[\omega_n(t+\tau)\omega_m(t)]$ and learn its parameters by maximum likelihood estimation using metered data. A convenient form is the so-termed \emph{Kronecker model}~\cite{KZG14}, according to which the covariance is expressed as the product of a spatial and a temporal kernel as $\mathbb{E}[\omega_n(t+\tau)\omega_m(t)] = k_\text{bus}(n,m)\cdot k_\text{time}(t+\tau,t)$. Our experimentation with various forms for $k_\text{bus}(n,m)$ and $k_\text{time}(t+\tau,t)$, including spatial covariances decaying with the electrical distance or propagation delays, was not particularly fruitful. This is justified because the Kronecker model is quite restrictive as it presumes that any pair of buses features the same time cross-covariance function. 

To arrive at an effective choice for $\mathbb{E}[\omega_n(t+\tau)\omega_m(t)]$, we should better encode any prior knowledge on the problem at hand. Given the complexity of grid dynamics, we proceed as:
\renewcommand{\labelenumi}{\emph{\alph{enumi})}}
\begin{enumerate}
\item Leverage an approximate stylized model of grid dynamics to decide a parametric form for $\mathbb{E}[\omega_n(t+\tau)\omega_m(t)]$.
\item Use metered data to estimate the parameters involved in the parametric form for $\mathbb{E}[\omega_n(t+\tau)\omega_m(t)]$. On the field, these data will be actual PMU readings. For the purposes of this work, data will be synthesized using realistic power system models. Heed that the models used to generate data in our numerical tests have not been derived by the stylized approximate dynamic grid model studied under \emph{a)}. The purpose of that model was only to provide reasonable parametric forms for covariances $\mathbb{E}[\omega_n(t+\tau)\omega_m(t)]$.
\item Plug metered data and the learned covariances in \eqref{eq:x2} to obtain point and uncertainty estimates for the non-metered signals of interest.
\end{enumerate}

The rest of this section targets \emph{a)} by reviewing and expanding upon an approximate model of grid dynamics. To obtain a simple yet sufficiently representative model, the key idea is to shift focus to an intrinsic set of system \emph{eigenstates}. Modeling eigenstates instead of the states $\omega_n$'s provides a physics-informed way to capture correlations across $\omega_n$'s, and thus, extrapolate across buses. Our workflow to obtain a parametric form for $\mathbb{E}[\omega_n(t+\tau)\omega_m(t)]$ involves three steps: \emph{S1)} Transform grid dynamics to a more convenient space of \emph{eigenstates}; \emph{S2)} Model eigenstates as GPs; and \emph{S3)} Convert eigenstates back to $\omega_n$'s. These steps are delineated next.

\emph{S1) Decoupling the MIMO Dynamical System:}
With $\bomega$ being the state, the transfer function of the system in \eqref{eq:swing} is
\begin{equation}\label{eq:tf1}
\bH(s)=s\left(s^2\bM+s\bD+\bL\right)^{-1}
\end{equation}
with $s$ being complex frequency of the Laplace domain. This transfer function simplifies significantly under the next assumption, which is adopted frequently to approximate power system dynamics~\cite{Low18},~\cite{Paganini19}.
	
\begin{assumption}[Uniform damping]\label{as:1}
The ratio of each generator's damping coefficient to its inertia is constant or $\bD=\gamma \bM$ for a given $\gamma >0$.
\end{assumption}

This assumption relies on the fact that both inertia and damping coefficients of a synchronous machine scale with the machine's power rating~\cite{Poolla17}. Under this assumption, the transfer function of swing dynamics can be rewritten as~\cite{Paganini19}
\[\bH(s)=s\bM^{-1/2}\bV\left(s^2\bI+s\gamma\bI+\bLambda\right)^{-1}\bV^\top\bM^{-1/2}\]
where $\bL_M=\bV \bLambda \bV^{\top}$ is the eigenvalue decomposition of matrix $\bL_M := \bM^{-1/2} \bL \bM^{-1/2}$. Because $\bL_M$ is psd, its eigenvalues have non-negative real values and are sorted in increasing order as $0=\lambda_1<\lambda_2\leq \ldots\leq \lambda_N$. These eigenvalues are placed on the main diagonal of matrix $\bLambda$. Moreover, the eigenvectors of $\bL_M$ are real-valued and orthonormal. They are placed as columns of $\bV$. 

Let us now transform the original inputs/states of \eqref{eq:swing} to the \emph{eigeninputs/eigenstates}~\cite{Zhu18},~\cite{Paganini19}
\begin{equation}\label{eq:trans}
\by(t) := \bV^{\top} \bM^{1/2} \boldsymbol{\theta}(t)\quad \text{and}\quad \bx(t):=\bV^{\top} \bM^{-1/2} \bp(t).
\end{equation}
Then the swing dynamics of \eqref{eq:swing} transform to
\begin{align}\label{eq:eigenspace}
\ddot{\by}(t) + \gamma \dot{\by}(t) + \bLambda \by(t) = \bx(t).
\end{align}
As $\bLambda$ is diagonal, the original MIMO system decouples into $N$ SISO \emph{eigensystems}. Eigensystem $i$ is described as
\begin{equation}\label{eq:diag}
\ddot{y}_i + \gamma \dot{y}_i + \lambda_i y_i = x_i
\end{equation}
which complies with the SISO example of \eqref{eq:siso}. If $\dot{y}_i(t)$ is selected as the system output (state), the impulse response of this system can be found to be~\cite{Zhu18}
\begin{equation}\label{eq:h}
    h_i(t)= \left(a_i e^{c_i t} + b_i e^{d_i t}\right) u(t)
\end{equation}
where $a_i,b_i:=\frac{1}{2} \mp \frac{\gamma}{2\sqrt{\gamma^2-4\lambda_i}}$; $c_i,d_i := \frac{-\gamma}{2} \pm \frac{\sqrt{\gamma^2-4\lambda_i}}{2}$; and $u(t)$ is the unit step function.

Thanks to this decoupling, we next propose a statistical model for eigenstates $\dot{y}_i$'s rather than speeds $\omega_n$'s. Although our modeling relies on Assumption~\ref{as:1}, it should be emphasized that our numerical tests of Section~\ref{sec:tests} were conducted on power networks not satisfying this assumption.



\emph{S2) Modeling Eigenstates as GPs:} To model eigenstates $\dot{\by}$, let us first study the eigeninputs $\bx(t)$. {Lacking specific information on the system inputs $\bp(t)$, we model them as random processes with non-informative prior distributions.} In particular, we postulate $\bp(t)$ to be a zero-mean white GP with covariance $\mathbb{E}[\bp(t+\tau)\bp^\top(t)]=\bSigma_p\delta(\tau)$, where $\delta(\tau)$ is the Dirac delta function. This means that the energy for each input $p_n(t)$ is equally distributed across frequencies. This way $\bx(t)$ is a zero-mean GP with covariance computed from~\eqref{eq:trans} as
\begin{equation}\label{eq:xcov}
\mathbb{E}[\bx(t+\tau)\bx^\top(t)]=\bA \delta(\tau)
\end{equation}
where $\bA:=\bV^{\top} \bM^{-1/2}\bSigma_p\bM^{-1/2}\bV$. It is natural to assume that $\bSigma_p$ is a diagonal matrix. In fact, if $\bSigma_p=\alpha\bM$ for some $\alpha>0$, matrix $\bA$ simplifies as $\bA=\alpha\bI_N$~\cite{Zhu18},~\cite{Zhu21},~\cite{Paganini19}; our prior work was limited to this special case~\cite{GPCDC}. A more reasonable model could be argued to be $\bSigma_p=\alpha\bM^2$ so that the standard deviation (and not the variance) of $p_n(t)$ scales with the inertia (and hence the power rating of bus $n$)~\cite{Paganini19}. To capture scenarios of any diagonal $\bSigma_p$ or even correlated inputs $\bp(t)$ due to renewable generation, we consider a general matrix $\bA$ in \eqref{eq:xcov} that is to be found. Having modeled the eigeninputs $\bx$, we proceed with the eigenstates $\dot{\by}$.

Consider the $i$-th eigensystem of \eqref{eq:diag}. It is known that when an LTI system is driven by a wide-sense stationary (WSS) random process $x_i(t)$, its output is a WSS random process too~\cite{leongarcia08}.\footnote{A random process $z(t)$ is WSS if its mean $\mathbb{E}[z(t)]$ and autocovariance function $\mathbb{E}[z(t+\tau)z(t)]$ do not depend on $t$~\cite[Ch.~9]{leongarcia08}.} The following proposition summarizes the statistical characterization of $\dot{y}_i(t)$; see also~\cite{Zhu18} for a related claim.

\begin{proposition}\label{pro:ki}
If the input $x_i(t)$ to the $i$-th eigensystem is a zero-mean white GP with variance $\alpha_{ii}$, the system output $\dot{y}_i(t)$ is a zero-mean GP with covariance 
\begin{equation}\label{eq:ki}
    \mathbb{E}[\dot{y}_i(t+\tau) \dot{y}_i(t)] = \alpha_{ii} k_{ii}(\tau)
\end{equation}
where $\alpha_{ii}$ is the $(i,i)$-th entry of $\bA$ defined in \eqref{eq:xcov}, and $k_{ii}(\tau):=\frac{1}{2\gamma}\left[h_i(\tau)+h_i(-\tau)\right]$.
\end{proposition}

\begin{IEEEproof}
As $x_i(t)$ is a zero-mean GP, the output $\dot{y}_i(t)$ is a zero-mean GP. Its covariance can be computed as~\cite[Ch.~10]{leongarcia08}
\begin{align*}
\mathbb{E}[\dot{y}_i(t+\tau)\dot{y}_i(t)]&= h_i(\tau)*h_i(-\tau)*\mathbb{E}[x_i(t+\tau)x_i(t)]\\
&=\alpha_{ii} h_i(\tau)*h_i(-\tau)
\end{align*}
The first equality holds for the output of any LTI system driven by a WSS random process. The second equality stems from the sifting property of the delta function. Computing the convolution establishes the claim. The result can alternatively be shown in the Laplace domain.
\end{IEEEproof}

Cross-covariances between eigenstates are found similarly.


\begin{proposition}\label{pro:kij}
If eigeninputs $\bx(t)$ are zero-mean white GPs with the covariance of \eqref{eq:xcov}, then
\begin{equation}\label{eq:Eyy2}
\mathbb{E}[\dot{y}_i(t+\tau) \dot{y}_j(t)] = \alpha_{ij} k_{ij}(\tau)u(\tau)+\alpha_{ij}k_{ji}(-\tau)u(-\tau)
\end{equation}
where $k_{ij}(\tau):=a_{ij} e^{c_i\tau} +b_{ij}e^{d_i \tau}$ and
\begin{align*}
a_{ij}&:=\frac{-c_i^2}{(c_i+c_j)(c_i+d_j)(c_i-d_i)}\\
b_{ij}&:=\frac{d_i^2}{(c_j+d_i)(d_i+d_j)(c_i-d_i)}
\end{align*}
and $\alpha_{ij}$ is the $(i,j)$-th entry of $\bA$ defined in \eqref{eq:xcov}.
\end{proposition}

\begin{IEEEproof}
The cross-covariance between the outputs of two LTI systems driven by WSS processes can be computed as
\begin{align*}
\mathbb{E}[\dot{y}_i(t+\tau)\dot{y}_j(t)]&= h_i(\tau)*h_j(-\tau)*\mathbb{E}[x_i(t+\tau)x_j(t)]\\
&=\alpha_{ij} h_i(\tau)*h_j(-\tau).
\end{align*}
The convolution $h_i(\tau)*h_j(-\tau)$ can be evaluated from tables as $k_{ij}(\tau)u(\tau)$ for $\tau\geq 0$. For $\tau<0$, it holds
\[\mathbb{E}[\dot{y}_i(t+\tau)\dot{y}_j(t)]=\mathbb{E}[\dot{y}_i(t)\dot{y}_j(t-\tau)]=\mathbb{E}[\dot{y}_j(t-\tau)\dot{y}_i(t)].\]
The last term equals $k_{ji}(-\tau)u(-\tau)$ because now $-\tau>0$.
\end{IEEEproof}

So far, we have computed the auto- and cross-covariances between eigenstates $\dot{y}_i$'s. The covariances between any pair of $(y_i,\dot{y}_i,\ddot{y}_i)$ and for any pair of $(i,j)$ can be computed by time integration or differentiation of the corresponding $k_{ij}(t_1,t_2)=k_{ij}(t_2+\tau,t_2)=k_{ij}(\tau)$ as discussed in Section~\ref{sec:GP}.  

\emph{S3) Modeling System States as GPs:} Having statistically modeled the eigenstates, it is now easy to model $\omega_n$'s. If $\dot{\by}(t)$ is a zero-mean GP, then $\bomega$ is also a zero-mean GP with covariance computed from \eqref{eq:trans} and \eqref{eq:ki}--\eqref{eq:Eyy2} as
\begin{equation}\label{eq:cov2}
\mathbb{E}[\bomega(t+\tau)\bomega^\top(t)]=\bM^{-1/2}\bV\mathbb{E}[\dot{\by}(t+\tau)\dot{\by}^\top(t)]\bV^\top\bM^{-1/2}.
\end{equation}

The covariance of \eqref{eq:cov2} completes our GP model and allows us to use the Bayesian framework of \eqref{eq:x2} to perform a broad range of learning tasks related to grid dynamics. This is because once we have a covariance (kernel) function for $\bomega$, we can easily compute covariances between $\btheta$, $\bomega$, $\dot{\bomega}$, and $\bp$, as detailed in the next Section~\ref{sec:inference}. Moreover, Section~\ref{sec:inference} explains how modeling states through eigenstates offers additional computational and observability advantages. Before that, we next consider a more detailed model of power system dynamics.

The swing dynamics of~\eqref{eq:swing} can be augmented to incorporate droop turbine control. For generator $n$, let $\tau_n$ be the time constant of its turbine and $r_n$ its droop control coefficient. The effect of the governor is ignored due to its relatively fast response compared to the turbine. Incorporating droop control, power system dynamics can be modeled as
		\begin{align*}
		\bM \dot{\bomega} +\bD \bomega + \bL \btheta &= \bp + \bp_{c}\\
		\bT \dot{\bp}_{c} + \bp_{c} &= -\bR^{-1} \bomega
		\end{align*}
where diagonal matrices $\bT$ and $\bR$ collect the turbine and droop parameters; and vector $\bp_c$ the adjustments in mechanical power. By eliminating $\bp_c$, we arrive at the third-order model
\begin{equation*}
\bT \bM \ddot{\bomega} + (\bM+\bT\bD)\dot{\bomega} + (\bD + \bR^{-1} + \bT\bL ) \bomega + \bL \btheta =\bT \dot{\bp}+ \bp
\end{equation*}
As in~\eqref{eq:swing}, the above MIMO system can be decomposed into $N$ SISO eigensystems under the ensuing assumptions~\cite{Paganini19}.

\begin{proposition}[\cite{Paganini19}]\label{prop:turbine}
Grid dynamics including turbine and droop control can be decomposed into $N$ SISO systems
\begin{equation*}
\tau \dddot{y}_i+(1+\gamma \tau)\ddot{y}_i+(\gamma+r+\tau \lambda_i)\dot{y}_i+\lambda_i y_i=\tau \dot{x}_i+x_i
\end{equation*}
through the transformation of~\eqref{eq:trans} if $\bD=\gamma \bM$, $\ \bR^{-1}=r\bM$, and $\bT=\tau\bI$ for given positive $\gamma$, $\tau$, and $r$. 
\end{proposition}


%

\section{Inferring Power System Dynamics}\label{sec:inference}

\begin{figure}[t]
\centering
\includegraphics[scale=0.5]{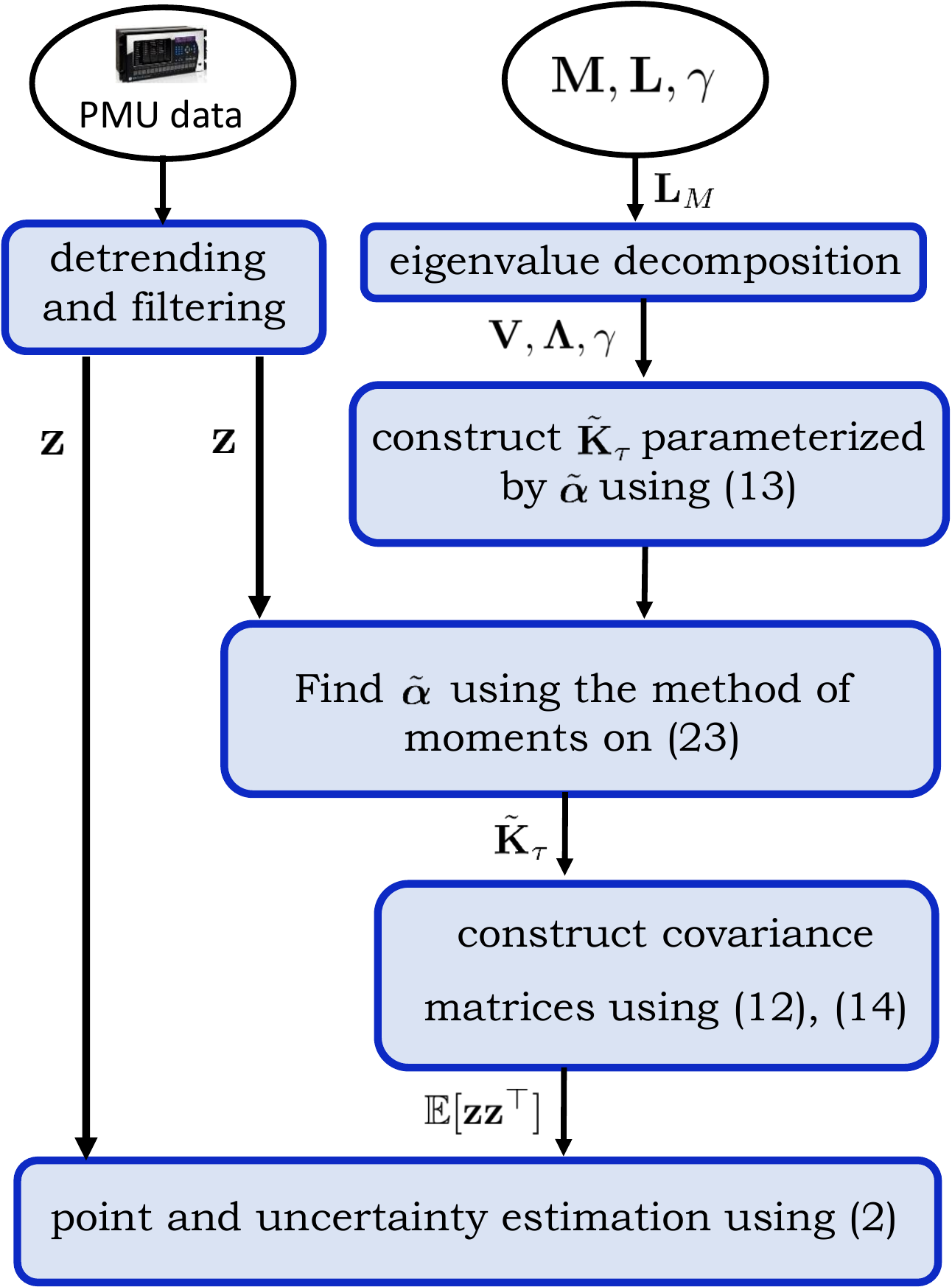}
\caption{{Basic steps for learning grid dynamics from synchrophasor data.}}
\label{fig:steps}  
\end{figure}

This section explains how to practically infer dynamic grid signals using the aforesaid GP-based framework. Before delving into the details, let us summarize the key steps of the process as illustrated in Fig.~\ref{fig:steps}. First, given matrix $\bL_M$, compute its eigenvalue decomposition to obtain matrices $\bV$ and $\bLambda$. System parameters (such as inertia coefficients, current operating point, line impedances, and network topology) are assumed known. If a specific range of oscillations is of interest (such as the range of inter-area oscillations), the operator may use only the say $D$ eigenvalues falling within that frequency range; we will elaborate on this feature later in Section~\ref{subsec:reduction}. Given $\gamma$ and $\{\lambda_i\}_{i=1}^D$, construct then the eigenstate covariance matrices from \eqref{eq:ki}--\eqref{eq:Eyy2}. At this stage, the covariance matrices are in parametric form since the values of $\alpha_{ij}$'s are not known. Upon filtering synchrophasor data, find $\alpha_{ij}$'s via the method of moments as explained in Section~\ref{subsec:mom}. Once $\alpha_{ij}$'s have been estimated, find the complete covariance of system oscillations using Proposition~\ref{pro:kij}. Having modeled the covariance between the collected measurements and the unknown dynamic grid signals of interest (see Section~\ref{subsec:infA} for details), provide point and uncertainty estimates for the latter using the Bayesian estimation formulas of~\eqref{eq:x2}. We next elaborate on the details.


\subsection{Spatiotemporal Covariance of Dynamic Oscillations}\label{subsec:infA}
To infer power system dynamics, we first need to identify which quantities are given [vector $\bx_1$ in \eqref{eq:jointGP}] and which are to be found [$\bx_2$ in \eqref{eq:jointGP}]. We then need to find their joint covariance. Knowing this matrix, the sought quantities and their uncertainties can be found from \eqref{eq:x2}.

Suppose the inference task refers to $T$ times comprising set $\mcT$. To simplify the exposition, we assume all measured signals are sampled for all $t\in\mcT$. Arrange speeds across \emph{all} buses and times in the $T\times N$ matrix 
\begin{equation}\label{eq:Theta}
\bOmega:=[\bomega(1)~\bomega(2)~\cdots~\bomega(T)]^\top.
\end{equation}
Stacking the columns of $\bOmega$ on top of each other yields the $TN$-long vector $\bomega:=\vectorize(\bOmega)$. The matrices of angles $\bTheta$, powers $\bP$, and eigenstates $\dot{\bY}$ are defined similarly. These matrices get vectorized into the $TN$-long vectors $(\btheta,\bp,\dot{\by})$, accordingly.

Define the $3TN$-long vector $\bz:=[\btheta^\top~~\bomega^\top~~\bp^\top]^\top$ collecting all random variables of interest; we ignore accelerations to keep the notation uncluttered. Measured and wanted quantities are subsets of the entries of $\bz$. For example, one may be measuring angles at a subset of buses $\mcM$ over $\mcT$, and would like to infer the angles or speeds at the remaining buses $\mcN\setminus \mcM$ again over $\mcT$. Or, one may be measuring angles and powers at some buses, and would like to find the speeds at all buses. All such scenarios can be handled using the formulas of \eqref{eq:x2} granted the related means and covariances are known.

To keep the exposition general, let us find the mean and covariance of the entire $\bz$. Lacking prior information, vector $\bz$ has been modeled as zero-mean. For its covariance, we start with the $\mathbb{E}[\bomega\bomega^\top]$ block of $\mathbb{E}[\bz\bz^\top]$. From \eqref{eq:trans} and \eqref{eq:Theta}, it follows that $\bOmega=\dot{\bY}\bV^\top\bM^{-1/2}$. Upon vectorizing both sides and using the property of the Kronecker product $\vectorize(\bA\bB\bC)=(\bC^\top\otimes \bA)\vectorize(\bB)$, we get that $\bomega=(\bM^{-1/2}\bV\otimes \bI_T)\dot{\by}$, so that
\begin{equation}\label{eq:cov3}
\mathbb{E}[\bomega\bomega^\top]=(\bM^{- 1/2}\bV\otimes \bI_T)\mathbb{E}[\dot{\by}\dot{\by}^\top] 
(\bV^\top\bM^{-1/2}\otimes \bI_T).
\end{equation}
Matrix $\mathbb{E}[\dot{\by}\dot{\by}^\top]$ has a block structure where the $(i,j)$-th $T\times T$ block stores $\mathbb{E}[\dot{y}_i(t)\dot{y}_j(t')]$ for all $t,t' \in \mcT$. These blocks can be computed from \eqref{eq:ki}--\eqref{eq:Eyy2}. The remaining blocks of the covariance $\mathbb{E}[\bz\bz^\top]$ can be found similarly. For example, the covariance $\mathbb{E}[\btheta\bomega^\top]$ can be modeled by replacing $\mathbb{E}[\dot{\by}\dot{\by}^\top]$ in~\eqref{eq:cov3} with $\mathbb{E}[\by\dot{\by}^\top]$ and integrating the covariances of \eqref{eq:ki}--\eqref{eq:Eyy2} with respect to their first time argument.


Depending on the learning setup, not all entries of $\bz$ are needed. If angles and powers are not measured or they are not to be inferred, then $\bz$ does not contain any entries from $\btheta$ or $\bp$. In particular, suppose we measure speeds at a subset of buses $\mcM$ and would like to find the speeds at bus $n\notin \mcM$. The part of $\bz$ corresponding to the collected data can be written as $\bz_\mcM=\vectorize(\bOmega\bS_\mcM)$ where matrix $\bS_\mcM$ selects the columns of $\bOmega$ related to $\mcM$. In this case, the required covariance matrix $\mathbb{E}[\bz\bz^\top]$ is obtained from \eqref{eq:cov3} by replacing $\bM^{-1/2}\bV$ with $\bS_\mcM^\top\bM^{-1/2}\bV$. Different bus selection matrices may have to be used for $\btheta$, $\bomega$, and $\bp$. Selection matrices may also be needed for sampling across time. For example, to collect angles over times $\mcT'\subseteq \mcT$, we can premultiply $\bOmega$ with a selection matrix $\bS_{\mcT'}$. This would mean replacing the identity matrix $\bI_T$ in \eqref{eq:cov3} with $\bS_{\mcT'}$. Despite being mundane, the formulas are straightforward to comprehend and code. 

We have hitherto assumed that the part of $\bz$ related to collected data is noise-free. In reality, grid dynamic data include measurement noise and modeling inaccuracies (e.g., higher-order dynamics or Assumption~\ref{as:1} being violated). To keep the statistical model tractable, we postulate that the collected data have been corrupted by additive zero-mean Gaussian noise that is independent across measurements and time. To incorporate the effect of noise on $\bz$, add the diagonal matrix of all noise variances on the block of $\mathbb{E}[\bz\bz^\top]$ related to the measured data [cf. submatrix $\bSigma_{11}$ in \eqref{eq:jointGP}]. In fact, this noise component ensures $\bSigma_{11}^{-1}$ exists for all learning scenarios. Nevertheless, inverting $\bSigma_{11}$ can be computationally challenging when more buses are metered over longer time intervals. We next discuss a model reduction technique.

\subsection{Model Reduction of Spatiotemporal Covariances}\label{subsec:reduction}
It can be argued that our primary goal has been to estimate eigenstates {as they will provide a simple parametric form for covariance $\mathbb{E}[\omega_n(t+\tau)\omega_m(t)]$ across actual states.} Once eigenstates $\dot{\by}$ have been estimated, the original states $\bomega$ can be computed as linear combinations of $\dot{\by}$ from \eqref{eq:trans}. Nonetheless, vectors $\bomega$ and $\dot{\by}$ are of the same dimension $N$, and hence, $\dot{\by}$ may be unobservable unless all buses are metered. To bypass this challenge, the idea here is to focus on a reduced number of eigenstates associated with \emph{inter-area oscillations}~\cite{Kundur94}. Industry experience and recent analytical studies~\cite{Paganini19}, \cite{Scaglione21}, reveal that inter-area oscillations occupy the lower frequency spectrum of dynamic grid signals $\btheta$, $\bomega$, and $\dot{\bomega}$. As the name suggests, such oscillations can be observed over larger geographical areas or even across the entire power system~\cite[Ch.~12]{Kundur94}. Different from intra-area oscillations which can be damped effectively by local controllers, inter-area oscillations are harder to control and are thus of particular interest~\cite{PSS92}. 

\begin{figure}[t] 
\centering
\includegraphics[scale=0.12]{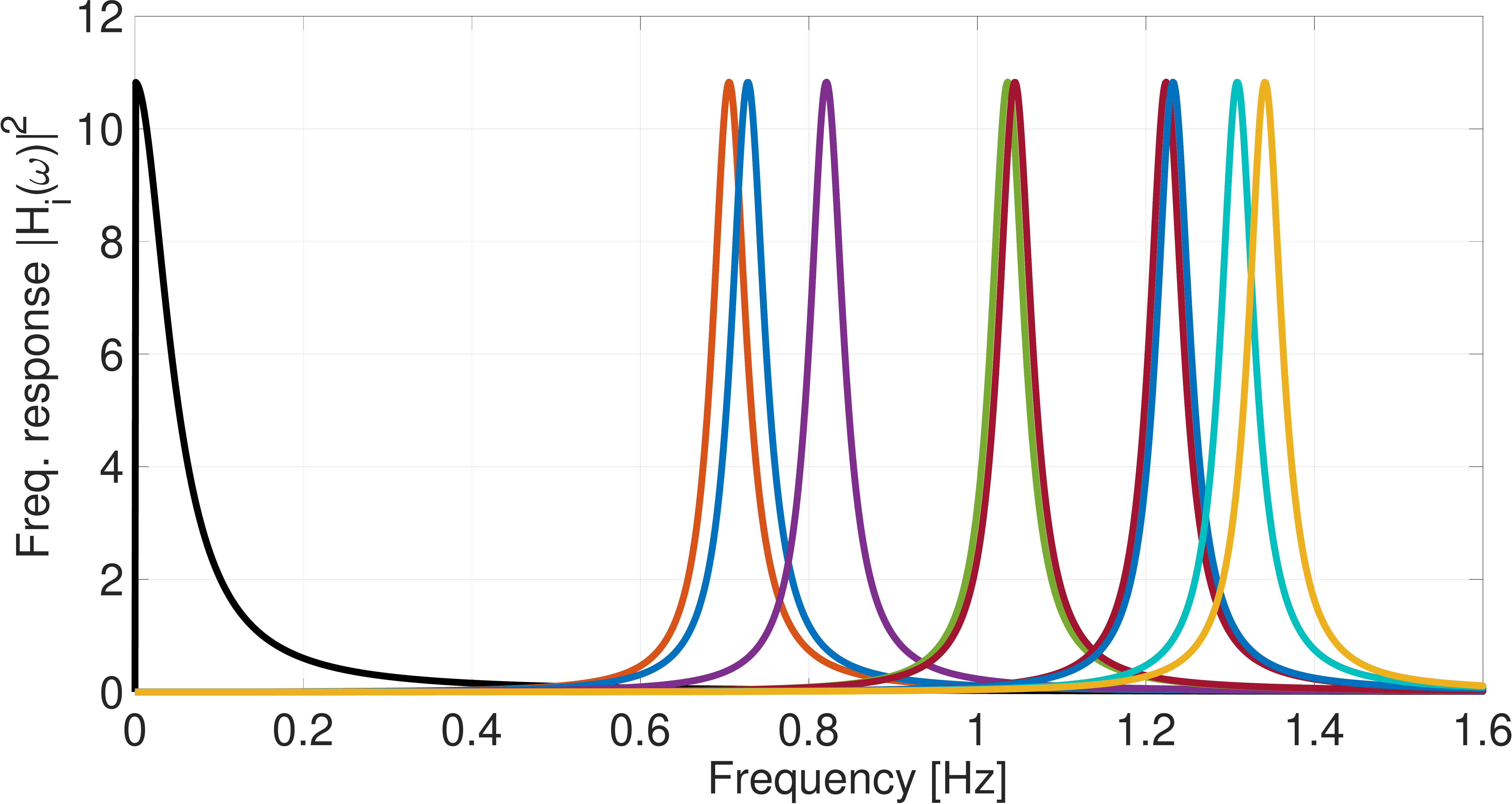}
\caption{Frequency responses of the first ten eigensystems for the IEEE 300-bus system.}
\label{fig:H}  
\end{figure}

To target inter-area oscillations, we leverage key frequency-domain properties of eigenstates. The frequency response of the $i$-th eigensystem can be computed from \eqref{eq:diag} as\footnote{We use the symbol $w$ for angular frequency of a signal to avoid confusion with the voltage speed at buses denoted by $\omega_n=\dot{\theta_n}$'s.}
\begin{equation}\label{eq:FR}
    |H_i(w)|^2=\frac{1}{(\frac{\lambda_i}{w}-w)^2+\gamma^2 }.
\end{equation}
Fig.~\ref{fig:H} plots $|H_i(w)|^2$ for the eigensystems associated with the ten smallest eigenvalues of matrix $\bL_M$ for the IEEE 300-bus power network. As evident from \eqref{eq:FR} and Fig.~\ref{fig:H}, each eigensystem $i$ exhibits a frequency selective behavior with its passband centered around the resonant frequency $w_i:=\sqrt{\lambda_i}$. Moreover, all eigensystems have the same frequency response value $H_i(w_i)=1/\gamma$ at their resonant frequency. Let us define the bandwidth of a system as the range of frequencies over which $|H_i(w)|^2$ is larger than half of its maximum value $1/\gamma^{2}$. Solving for $|H_i(w)|^2 = \frac{1}{2\gamma^2}$ yields the cutoff frequencies of eigensystem $i$:
\begin{equation*}
    \underline{w}_i := \frac{ - \gamma + \sqrt{\gamma^2+4\lambda_i}}{2}\quad\text{and} \quad \overline{w}_i:= \frac{ \gamma + \sqrt{\gamma^2+4\lambda_i}}{2}.
\end{equation*}
Then, the bandwidth of eigensystem $i$ is $\overline{w}_i -\underline{w}_i =\gamma$, which interestingly remains the same for all $i$. In a nutshell, eigensystems exhibit the same passband shape centered at resonant frequencies ordered as $0=w_1<w_2\leq \ldots\leq w_N$.

\begin{remark}\label{re:reminder}
Note that the aforesaid properties hold for the linearized swing dynamics and only under Assumption~\ref{as:1}. It is only then that the MIMO system of grid dynamics decouples into the set of $N$ SISO eigensystems. Recall that this stylized model of grid dynamics is used only to justify a reasonably complex yet representative prior in terms of a parametric form for $\mathbb{E}[\omega_n(t+\tau)\omega_m(t)]$. Once that form has been derived, the estimation of the covariance parameters and the inference of non-metered signals operate on on actual dynamic grid data, which do not fully comply with the stylized model. 
\end{remark}

\begin{figure}[t] 
\centering
\includegraphics[scale=0.22]{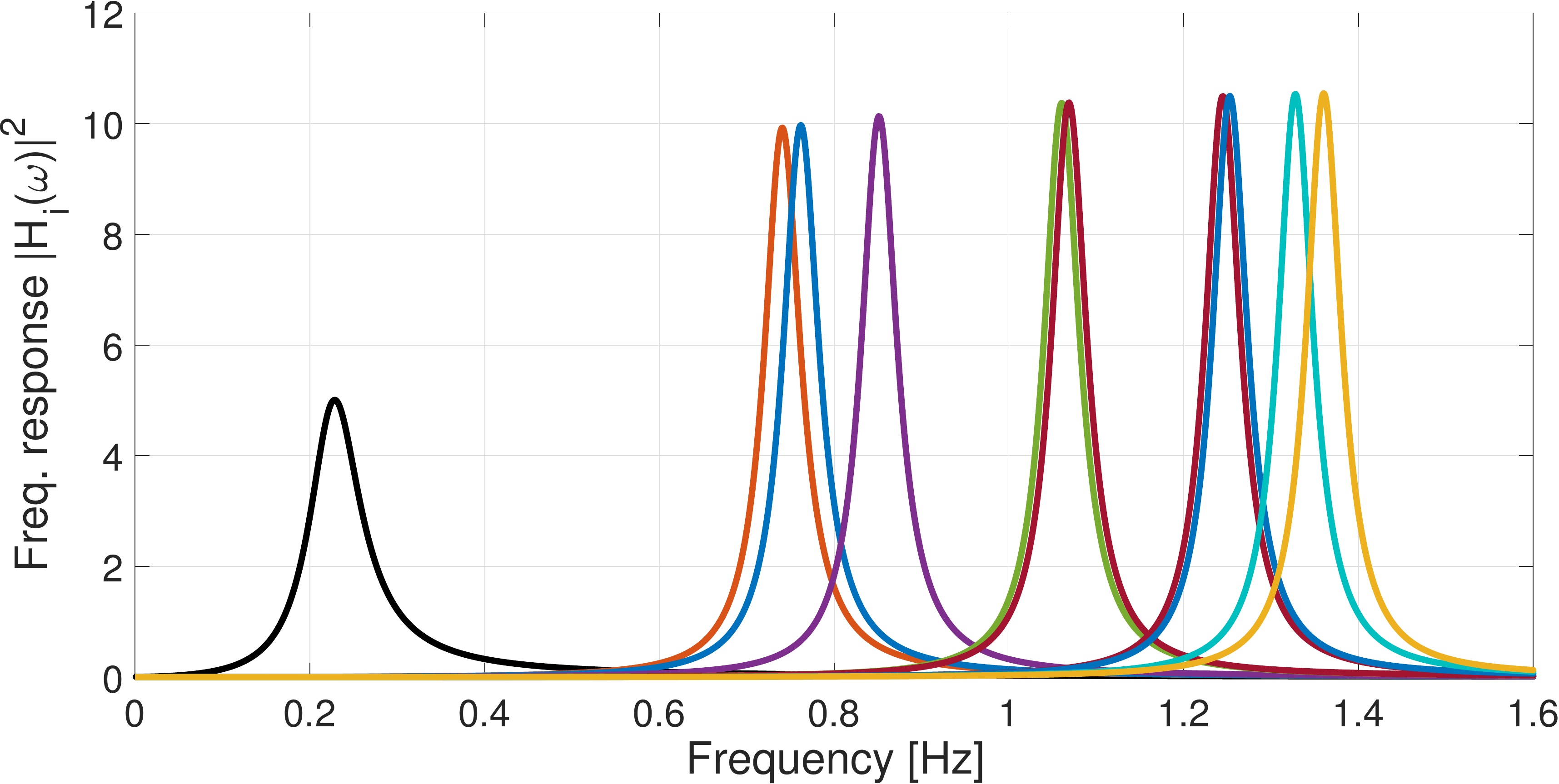}
\caption{{Frequency responses of the first ten eigensystems for the IEEE 300-bus system with turbine dynamics and droop control.}}
\label{fig:H_TG}  
\end{figure}

Due to the frequency selective shape of eigensystem $i$, the frequency content of eigenstate $\dot{y}_i(t)$ can be approximately confined within $(\underline{w}_i,\overline{w}_i)$. Hence $\dot{y}_i(t)$ can be associated to the resonant frequency $w_i$. Therefore, to study inter-area oscillations typically observed within the frequency range
\[\Omega:=[0.1,0.8]~\text{Hz},\]
the operator can confine its attention only on the eigenstates falling within that range. To explain this, let us partition $\dot{\by}$ into the subvector $\dot{\by}_1$ collecting all eigenstates falling within $\Omega$, and subvector $\dot{\by}_2$ collecting the remaining eigenstates. We partition the matrix of eigenvectors $\bV$ similarly into $\bV_1$ and $\bV_2$. Then, the transformation in \eqref{eq:trans} can be expanded as
\[\bomega=\bM^{-1/2}\bV\dot{\by}=\bM^{-1/2}\bV_1\dot{\by}_1+\bM^{-1/2}\bV_2\dot{\by}_2.\]
Interestingly, due to linearity, the frequency content of $\bomega(t)$ along $\Omega$ is attributed only to the first summand of the previous expansion. The second summand has no (or negligible) content along frequency range $\Omega$. We would like to estimate $\dot{\by}_1$, but the metered entries of $\bomega$ depend on both $\dot{\by}_1$ and $\dot{\by}_2$. Nonetheless, thanks to the separation of $\dot{\by}_1$ and $\dot{\by}_2$ in the frequency domain, we can tune out the effect of $\dot{\by}_2$ without knowing either $\dot{\by}_1$ or $\dot{\by}_2$. This can be achieved by simply passing rotor speed measurements through a filter having $\Omega$ as its pass band. If $\bomega_1(t)$ denotes the filtered $\bomega(t)$, we obtain that
\begin{equation}\label{eq:ftrans}
\bomega_1(t)= \bM^{-1/2}\bV_1 \dot{\by}_1(t).
\end{equation}

The importance of \eqref{eq:ftrans} is that without knowing the eigenstates $\dot{\by}_2$, their contribution can be removed from all measurements. Reference~\cite{Scaglione21} introduced this idea for graph signal processing of synchrophasor data. We can now use filtered speed data to estimate the inter-area eigenstates $\dot{\by}_1(t)$. Using \eqref{eq:ftrans}, we can replace \eqref{eq:cov3} by
\begin{equation}\label{eq:cov4}
\mathbb{E}[\bomega_1\bomega_1^\top]=(\bM^{- 1/2}\bV_1\otimes \bI_T)\mathbb{E}[\dot{\by}_1\dot{\by}_1^\top] 
(\bV_1^\top\bM^{-1/2}\otimes \bI_T).
\end{equation}

Aiming for $\dot{\by}_1$ rather than ${\tby}$ emphasizes on inter-area oscillations, is expected to improve upon observability, as well as offers computational benefits. To see the latter, suppose we maintain only $D\ll N$ eigenstates. We have replaced the $NT\times NT$ matrix $\mathbb{E}[\dot{\by}\dot{\by}^\top]$ of \eqref{eq:cov3} with the $DT\times DT$ matrix $\mathbb{E}[\dot{\by}\dot{\by}^\top]$ of \eqref{eq:cov4}. The computational gain is not only in the matrix products of \eqref{eq:cov4}, but also in the matrix inversion of $\bSigma_{11}$ in \eqref{eq:x2}. Matrix $\bSigma_{11}$ takes the form $\bSigma_{11}=\bB\mathbb{E}[\dot{\by}\dot{\by}^\top]\bB^\top+\sigma_n^2\bI$, where $\bB:=\bS_\mcM^\top\bM^{-1/2}\bV_1\otimes \bI_T$ is of dimension $MT\times DT$ if $M$ buses are metered. One can utilize the matrix inversion lemma to find $\bSigma_{11}^{-1}$ in $O(MD^2T^3)$ instead of $O(M^3T^3)$ operations with $D\ll M$. Focusing on inter-area oscillations and the bandpass nature of eigensystems simplify also the task of finding the parameters in \eqref{eq:ki}--\eqref{eq:Eyy2} as discussed next.

\begin{remark}\label{re:obs}
From~\eqref{eq:ftrans}, it holds that $\tilde{\omega}_n(t) = \be_n^\top \bM^{-1/2} \bV_1 \tby(t)$. Evidently, the $(n,i)$-th entry of $\bM^{-1/2}\bV_1$ determines the participation of $\dot{\tilde{y}}_i(t)$ in $\tilde{\omega}_n(t)$. 
Consider a scenario where the $i$-th eigenstate has negligible participation in all measured quantities. Naturally, the GP estimates of speeds at non-metered buses where $\dot{\by}_1$ has a high contribution will be inaccurate. 
\end{remark}


%

\subsection{Parameter Estimation}\label{subsec:mom}
To evaluate the kernel functions of \eqref{eq:ki}--\eqref{eq:Eyy2}, we need to know the $N^2$ entries of $\bA=\mathbb{E}[\bx(t)\bx^\top(t)]$. We use data to learn these entries $\alpha_{ij}$'s. If the focus is on $D$ inter-area eigenstates, we only need to find the corresponding $D\times D$ submatrix $\tbA$ of $\bA$. Parameters involved in GP models are typically found via maximum likelihood estimation~\cite{GP}, \cite{Bishop}, \cite{raissi17}, though such approaches are computationally complex and entail solving non-convex optimization problems. We propose a scalable estimation approach instead based on the \emph{method of moments} (MoM)~\cite[Ch.9]{Kay93}. 

Suppose the collected filtered data consist of speeds measured at a subset $\mcM$ of $M$ buses, that is $\bz(t)=\bS_\mcM\bomega_1(t)$. To find $\tbA$, we consider \eqref{eq:cov2} for lag $\tau$. With a slight abuse in notation, we define the $D\times D$ matrix $\tbK_\tau$ such that $[\tbK_\tau]_{ij}:=k_{ij}(\tau)$ for $i,j$ indexing the eigenstates within the frequency band of interest. Ignoring noise for now, it holds
\begin{equation}\label{eq:mom1}
\mathbb{E}[\bz(t+\tau) \bz^\top(t)]=\bS_{\mcM} \bM^{-1/2} \bV_1~\left( \tbA \odot \tbK_\tau \right)~\bV_1^\top \bM^{-1/2} \bS^\top_{\mcM}
\end{equation}
where $\odot$ denotes the element-wise product between two matrices. Note that the data covariance matrix depends linearly on $\tbA$. Matrix $\tbK_\tau$ is known from~\eqref{eq:Eyy2}. The ensemble covariance of \eqref{eq:mom1} can be approximated by the sample covariance matrix
\begin{equation}\label{eq:mom2}
\bC_\tau=\frac{1}{T}\sum_{t=1}^{T}\bz(t+\tau) \bz(t)^\top.
\end{equation} 
The MoM suggests using the sample in lieu of the ensemble covariance to estimate $\tbA$. Vectorizing \eqref{eq:mom1} and the Kronecker product property $\vectorize(\bA\bB\bC)=(\bC^\top\otimes \bA)\vectorize(\bB)$ yield that
\begin{equation}\label{eq:alpha}
\bc_\tau= \bU \vectorize(\tbA \odot \tbK_\tau )= \bU \diag(\tbk_\tau) \tbalpha
\end{equation}
 where $\bc_\tau:=\vectorize(\bC_\tau)$; $\tbalpha:=\vectorize(\tbA)$; $\tbk_\tau:=\vectorize(\tbK_\tau)$; and $\bU:=(\bS_{\mcM} \bM^{-1/2} \bV_1)\otimes(\bS_{\mcM} \bM^{-1/2} \bV_1)$. 

Due to the symmetry of $\bC_\tau$ and $\tbA$, we have to estimate $(D^2+D)/2$ parameters collected in $\tbalpha$. Exploiting the frequency selective nature of eigensystems, we can reduce further the number of $\alpha_{ij}$'s that need to be found: The eigeninputs $x_i(t)$ and $x_j(t)$ pass through eigensystems of known passbands. If the two passbands do not overlap, the corresponding entry of $\tbk_\tau$ is close to zero for all $\tau$, and so $\alpha_{ij}$ is irrelevant for our learning. In other words, this particular $\alpha_{ij}$ does not have to be found. We estimate the remaining entries of $\balpha$ or equivalently $\tbA$ using a least-squares (LS) fit on the entries of $\bc_0$ based on the linear model of \eqref{eq:alpha} subject to the constraint that the complete $\tbA$ is positive semidefinite.

To account for noise, an additional term $\diag(\bsigma)$ should be added on the right-hand side of \eqref{eq:mom1}. The $M$-length vector $\bsigma$ collects the variances for all terms of measurement noise. Assuming noise is white, the term appears only for $\tau=0$. If $\bsigma$ is known from the manufacturer of the metering devices or historical data, matrix $\diag(\bsigma)$ should be subtracted from $\bC_\tau$. Otherwise, vector $\bsigma$ can be estimated jointly with $\balpha$.

\subsection{Comparison with Existing Works}\label{subsec:comparison}
Having presented our GP-based methodology for learning grid dynamics, we are now able to compare it with prior works related to processing synchrophasor data. As reviewed in the Introduction, such works can be grouped into those using synchrophasor data alone and those falling under the dynamic state estimation (DSE) paradigm. With reference to the applications depicted in Fig.~\ref{fig:apps}, Table~\ref{tbl:comp} compares how each group of approaches handles each application and under what requirements. Data-based methods are pertinent when a power system model is not available, and the goal is to impute entries from a few datastreams that are otherwise metered. They can also be used for anomaly detection or locating oscillatory sources by triangularization. DSE methods on the other hand, build upon a detailed dynamic power system model to estimate, predict, or differentiate the unknown system state. They usually focus on a single bus or a single control area, and they presume that all inputs to this dynamical system are metered. Our proposed GP-based methodology applies to grid-wide inference tasks when observability criteria are not met, none or a few inputs are metered, and one can only rely on an approximate dynamic grid model. In addition to multi-datastream applications, the GP-based methodology can handle single-datastream applications of practical interest, such as finding voltage speeds from angles (differentiation), prediction, or imputation.

\begin{table}
	\renewcommand{\arraystretch}{1.1}
	\caption{Comparing Methods for Learning from PMUs per Application. (applications are defined in Figure~\ref{fig:apps})}
	\label{tbl:comp} 
	\centering
\begin{tabular}{l|p{2cm}|p{2.2cm}|p{2.2cm}}\hline\hline
&\makecell{Data-based\\\cite{chow14}--\cite{Chow20_tensor}}&\makecell{DSE\\\cite{DSE19}--\cite{Mili18}}&\makecell{GP approach}\\\hline\hline
\emph{A1)} & multi-stream & \checkmark * &\checkmark\\\hline
\emph{A2)} &\xmark & \checkmark * & \checkmark\\\hline
\emph{A3)} &\xmark & \checkmark * & \checkmark\\\hline
\emph{A4)} & multi-stream & \checkmark * & \checkmark\\\hline
\emph{A5)} &\xmark & \xmark & \checkmark\\\hline
\emph{A6)} & \cite{Liu_ISGT19}--\cite{le_xie20} & \xmark & \checkmark\\\hline
&No need for dynamic system input or model& 
Need system inputs and model (single bus/area) &
--\emph{Single-PMU applications} do not need inputs or model.\\
&&&--\emph{Multi-PMU applications} need $\gamma$, $\bL$, and $\bM$; inputs are not necessary\\
\hline\hline
\end{tabular}
\end{table}
 
\color{black}


\section{Numerical Tests}\label{sec:tests}

The novel GP-based learning framework was tested on several of the applications discussed at the end of Section~\ref{sec:GP}. Before elaborating on the applications, let us first explain how dynamic grid data were generated. Numerical tests were not conducted under the stylized dynamic grid model of Section~\ref{sec:power}, but on more realistic setups. In particular, data were synthesized using the IEEE 39- and 300-bus benchmarks. Upon Kron reduction, the two benchmarks were converted to 10- and 69-machine networks~\cite{Ishi18}. The IEEE 39-bus benchmark is also known as the 10-machine New England power system and has been widely used in the literature on dynamics. We used three models to simulate grid dynamics:

\begin{itemize}
    \item Model \emph{M1)} simulates the 69-machine system based on the linearized model of~\eqref{eq:swing}. Linearization was performed at the nominal operating point.
    \item Model \emph{M2)} is used for modeling the nonlinear 10-machine system using fourth-order generator models with automatic voltage regulators (AVR) included.
    \item Model \emph{M3)} models the nonlinear 10-machine system using fourth-order generator models along with turbine/governor dynamics, droop control, and AVR.
\end{itemize}

Both benchmarks violated Assumption~\ref{as:1} on constant damping-to-inertia coefficient ratios $D_n/M_n=\gamma$ for all $n$. In the 69-machine system, ratios lie within the relatively narrow range of $[0.19,0.4]$, so Assumption~\ref{as:1} may be reasonable. In the 10-machine benchmark however, $D_n/M_n$ ratios take on the distinct values $\{0.002,0.024,0.028,0.04\}$. In the same system, the ratios $r_n^{-1}/M_n$ of inverse-droop to inertia coefficients take values in $[0.09,6.8]$. Turbine time constants $\tau_n$'s take the distinct values of $\{8,8.7,10\}$~seconds. Even though $\tau_n$'s seem to be relatively close to each other, the $D_n/M_n$ and $r_n^{-1}/M_n$ ratios vary by one or two orders of magnitude, thus severely violating the assumptions stated in Assumption~\ref{as:1} and Proposition~\ref{prop:turbine}. Note that in the IEEE 39-bus benchmark, generator 1 represents an aggregation of a large number of generators, and that may explain the large disparity in ratios compared to the IEEE 300-bus (69-machine) benchmark. Either way, this 10-machine system serves as an excellent worst-case benchmark to evaluate the robustness of the proposed GP-based framework against discrepancies between the actual dynamic grid model and the stylized model used to derive covariances in Section~\ref{sec:power}. 

To derive the parametric form for covariances under models \emph{M1)} and \emph{M2)}, we need to know $(\bM,\gamma,\bL)$. Under \emph{M3)}, we additionally need to know $\tau$ and $r$. We chose
\[\gamma=\frac{\bone^\top \bD \bone}{\bone^\top \bM \bone},\quad r = \frac{\bone^\top \bR^{-1} \bone}{\bone^\top \bM \bone}, \quad \text{and}\quad \tau:=\frac{\bone^\top \bT\bone}{N}.\]
For \emph{M1)}, we obtained $\bL$ from MATPOWER's function \texttt{makeJac}. To generate data under \emph{M1)}, the swing equation was converted to its state-space representation using MATLAB's \texttt{ss}, and was solved using MATLAB's \texttt{ode45} for a given input $\bp(t)$. For \emph{M2)} and \emph{M3)}, we used the power system analysis toolbox PSAT~\cite{psat}, which further simulates automatic voltage regulators, turbine/droop control, and computes synchrophasor data to be used as PMU readings after adding measurement noise. 

We simulated both ambient and non-ambient (fault) conditions. Ambient conditions were tested under \emph{M1)} and \emph{M2)}, with power injections $\bp(t)$ being synthesized as zero-mean Gaussian processes having covariance $\mathbb{E}[\bp(t+\tau)\bp^\top(t)]=0.01 \bM^2 \delta(\tau)$. For non-ambient conditions, we simulated a fault scenario as follows: A three-phase fault occurred at a substation connected to one of the generators. The fault caused the breakers of adjacent lines to open after $3$~ms from the occurrence of the fault. After an additional $3$~ms, the breakers re-closed and the fault was cleared. Under \emph{M1)}, the described disturbance was simulated using an impulse function. Under \emph{M2)} and \emph{M3)}, PSAT simulated the fault and breaker operation in detail. Observe that under either ambient and fault conditions, the assumption on independent eigeninputs postulated by the stylized dynamic grid model of Section~\ref{sec:power} was not met.

Data were corrupted by zero-mean additive white Gaussian noise of variance $0.01$~rad for angle, and $0.005$~rad/sec for rotor speed data. Dynamics were simulated at a time resolution of $1$~ms, but processed at the reporting rate of $15$ samples per second~\cite{C37118}. Focusing on inter-area oscillations, measurements were passed through a band-pass filter with $[0.5,0.8]$~Hz passband using Matlab's \texttt{designfilt} and~\texttt{filtfilt} functions. Even though inter-area oscillations usually appear in the range of $[0.1,0.8]$~Hz, we selected a narrower passband as the adopted benchmarks did not exhibit resonant frequencies in $[0.1,0.5]$~Hz. The parameter matrix $\tilde{\bA}$ was estimated using the MoM. The $\alpha_{ij}$'s for eigensystems that have no overlap within $(2\underline{w}_i, 2\overline{w}_i)$ were set to zero.

The rest of this section is organized into four application setups: Section~\ref{subse:i} infers angle or speed oscillations at non-metered buses using angle or speed data at metered data under ambient and fault conditions. Section~\ref{subsection:ii} aims at locating a fault by estimating the power injections at all buses using rotor speed measurements. Sections~\ref{subsec:iii} and~\ref{subsec:iv} focus on inference problems related to a single bus, and hence, no dynamic grid model needed as covariances are now only functions of time and not space. Section~\ref{subsec:iii} imputes missing data from a single synchrophasor datastream, while Section~\ref{subsec:iv} estimates a rotor speed signal from a rotor angle datastream (time differentiation).

\subsection{Inferring Rotor Angles and Speeds at Non-metered Buses}\label{subse:i}
\subsubsection{Inferring Rotor Angles and Speeds Using M1)}

\begin{figure}[t]
\centering
\includegraphics[scale=0.22]{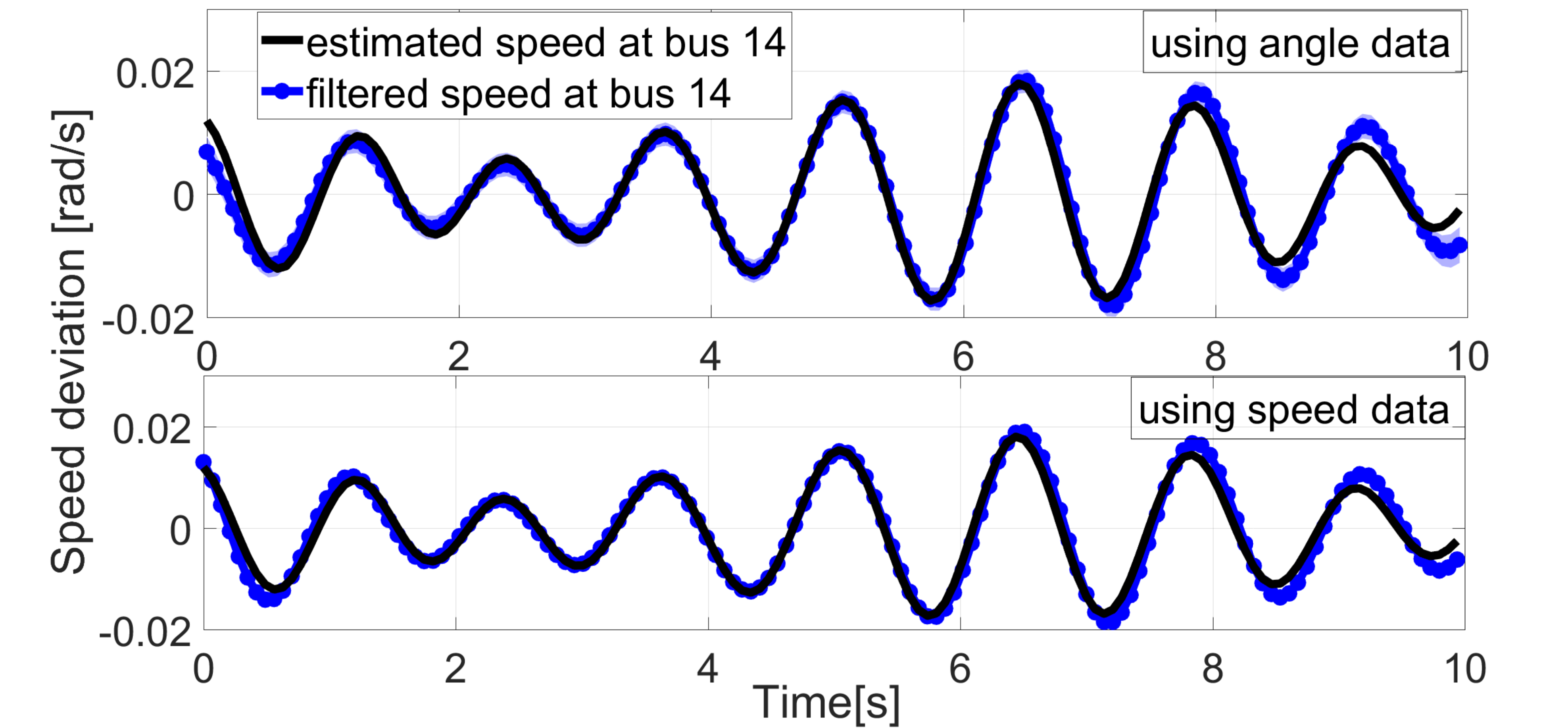}
\caption{Speed estimation at bus $14$ of the 69-machine system under ambient disturbances using angle (top panel) and speed (bottom panel) measurements. The shaded area shows the $\pm\sigma$ uncertainty.}	
\label{fig:L_omega}  
\end{figure}

We first tested the proposed method under \emph{M1)} excited by ambient disturbances. Measurements (angle or speed) were randomly collected from $50$ out of the $69$ generators. Angle measurements were contaminated by an additive white Gaussian noise with a standard deviation of $0.005$~rad to model noise under small-signal disturbances. We estimated speeds at non-metered buses using angle measurements first and speed measurements secondly. Fig.~\ref{fig:L_omega} depicts the obtained GP estimates. The shaded areas demonstrate the $\pm \sigma$ uncertainty interval obtained by taking the square root of the diagonal entries of the covariance matrix of~\eqref{eq:x2:c}. The results in Fig.~\ref{fig:L_omega} confirm the ability of our method to estimate speeds using angle or speed data. 

\begin{figure}[t]
	\centering
	\includegraphics[scale=0.24]{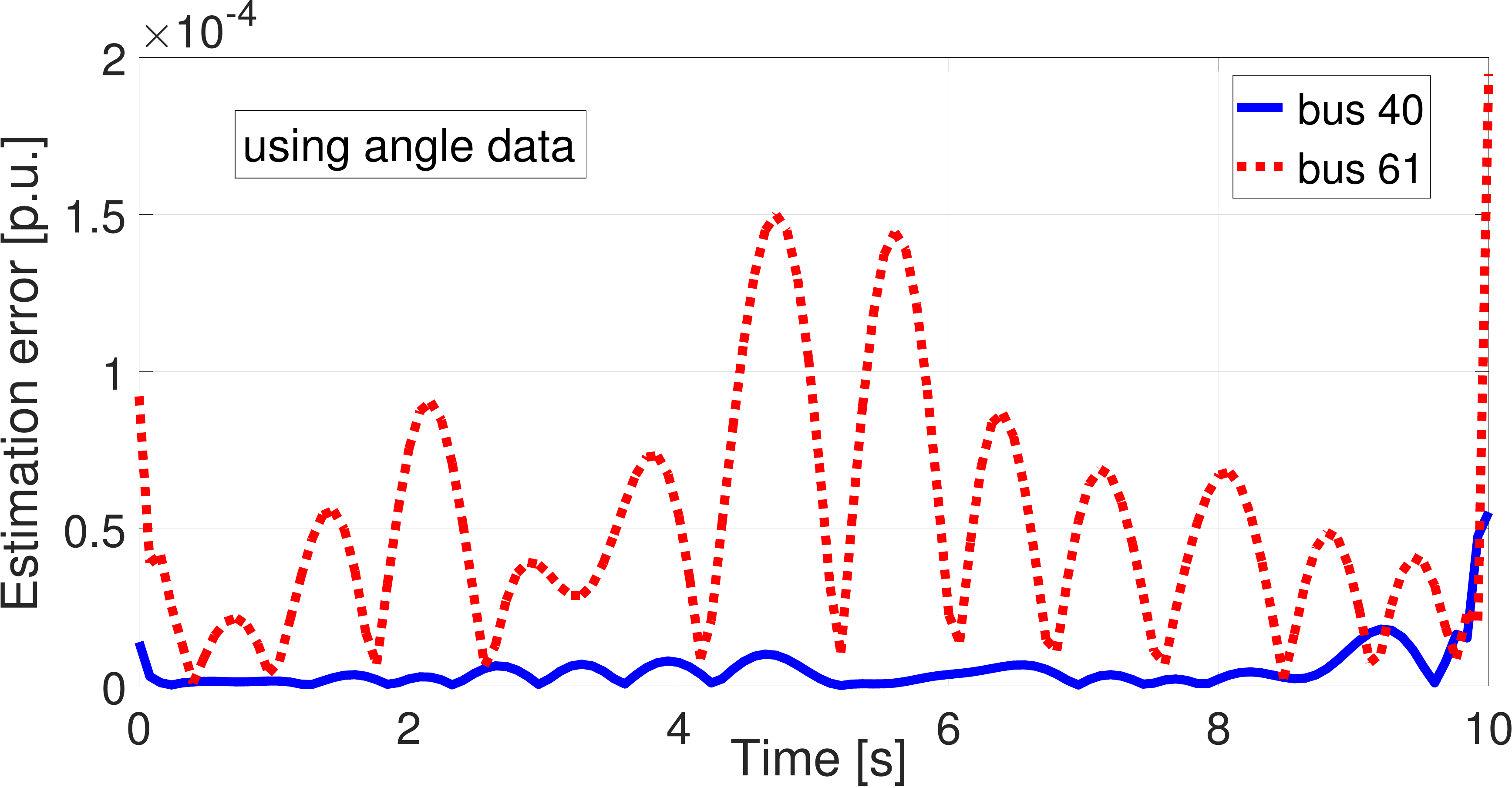}
	\includegraphics[scale=0.24]{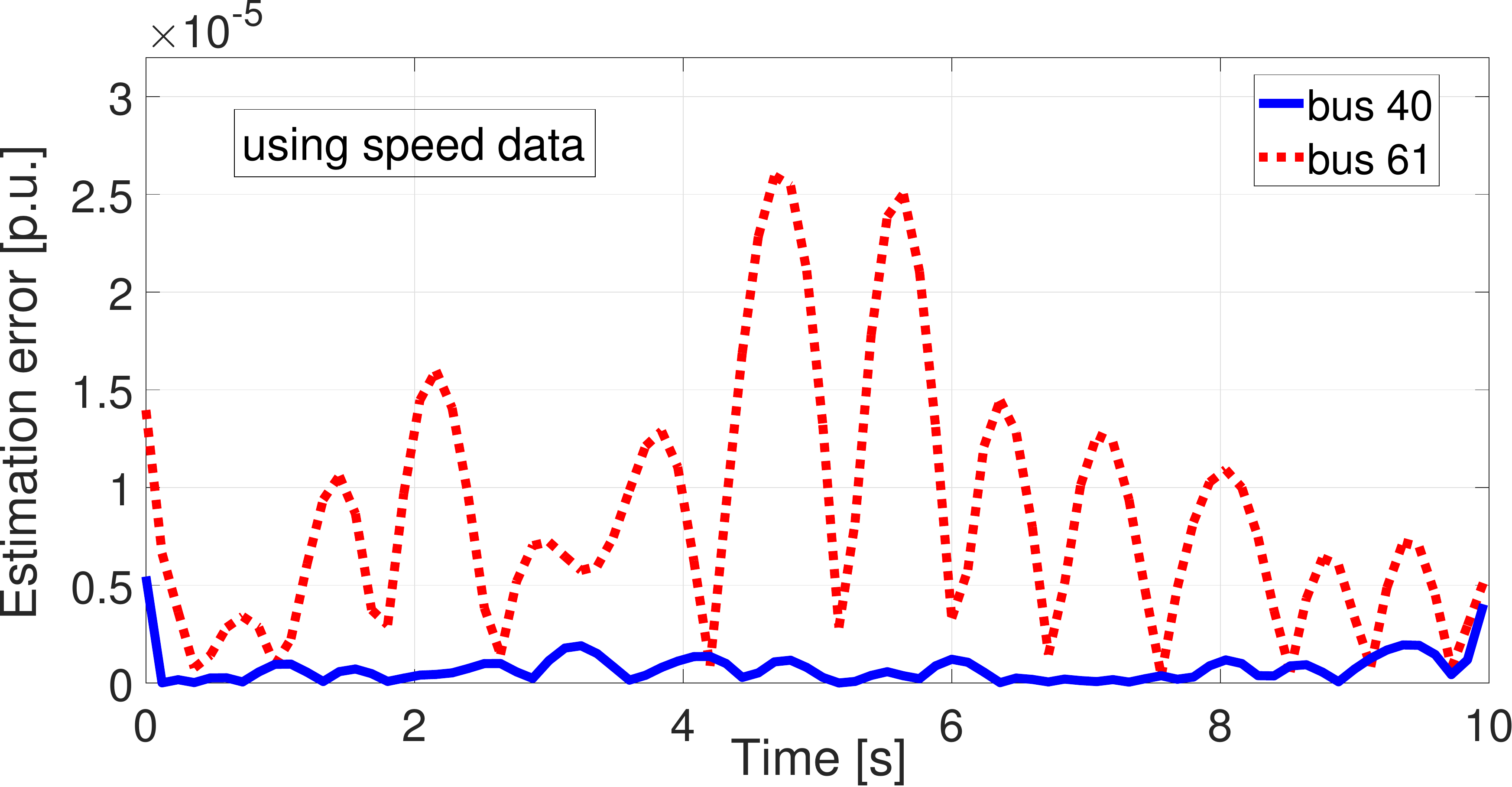}
	\caption{{Estimating rotor speeds under ambient dynamics on the 69-machine network using angle (top) and speed (bottom) data from 50 buses. Regardless of using angle or speed data, the buses with the largest and smallest time-averaged errors did not change.}}	
	\label{fig:error_fwt}  
\end{figure}

Estimation accuracy was measured as the absolute error $E_n(t)$ in per unit between the actual and the predicted speed at bus $n$ and time $t$
\begin{equation*}
    E_n(t) := \frac{|\omega_n(t) - \hat{\omega}_n(t)|}{\omega_0}\quad \text{and} \quad E_n :=\frac{1}{T} \sum_{t=1}^T E_n(t)
\end{equation*}
where $\omega_0$ is the nominal angular velocity. The error $E_n(t)$ for the buses attaining the largest and smallest time-averaged errors $E_n$ are shown in Fig.~\ref{fig:error_fwt}. The goal of this test was to estimate speeds at non-metered buses using angle or speed measurements at 50 randomly sampled buses. The error varies with time because the measurement noise variance remains constant across time while the actual signals oscillate. 

\begin{figure}[t]
	\centering
	\includegraphics[scale=0.22]{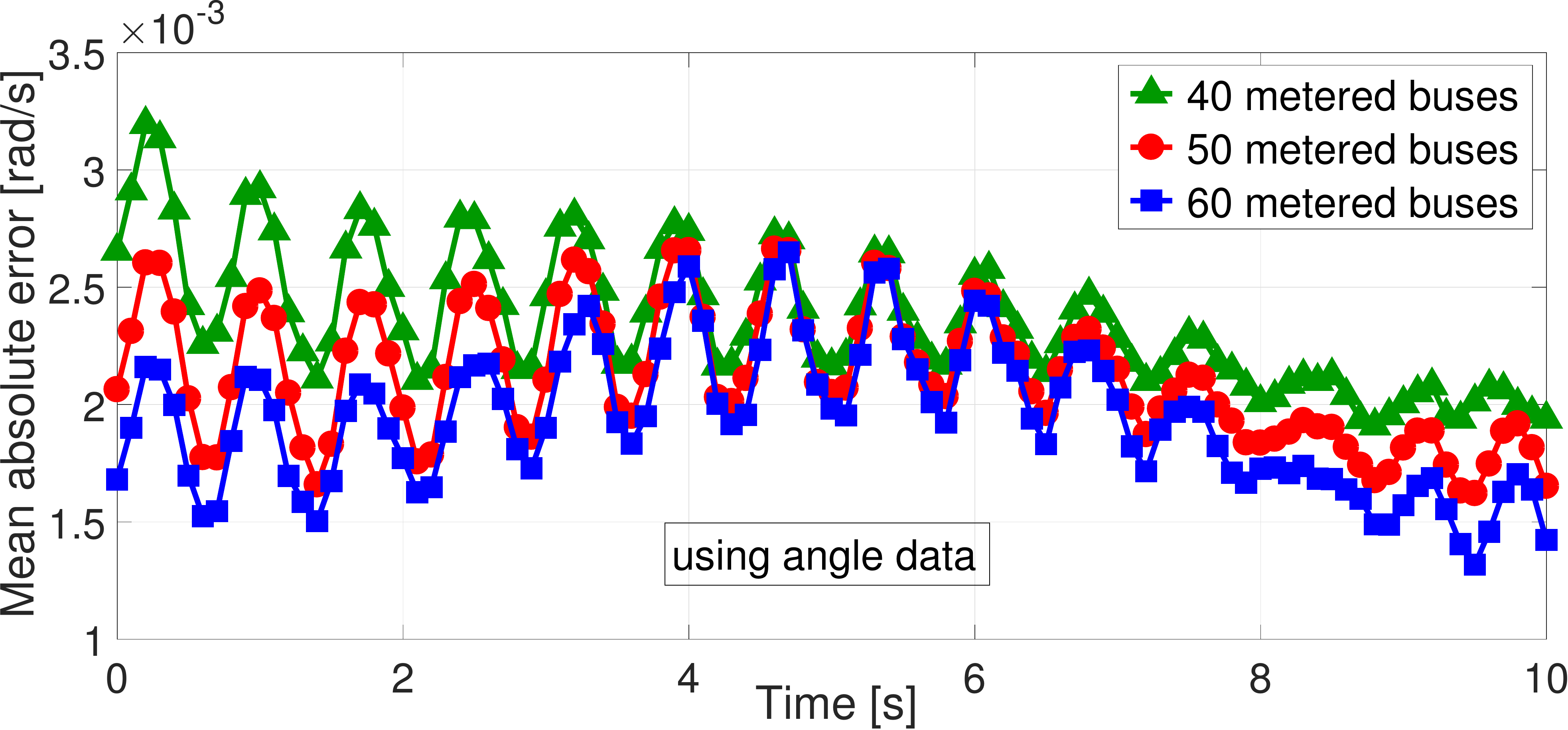}
	\includegraphics[scale=0.22]{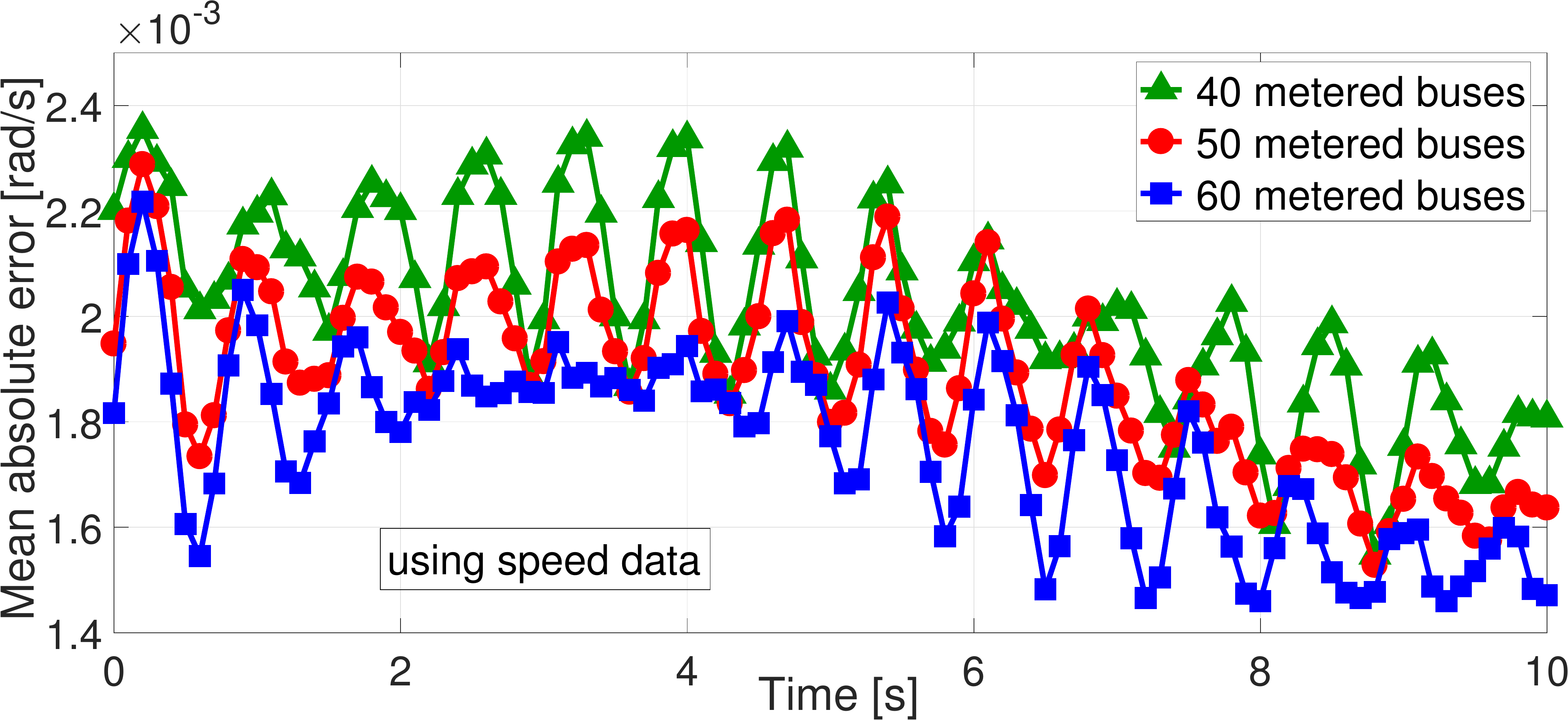}
	\caption{{Studying the effect of the number and placement of metered buses on estimating speeds from angle (top) and speed (bottom) data. Absolute errors on speeds have been averaged over the non-metered buses and over $100$ random metering placements of $40$, $50$, and $60$ buses.}}
	\label{fig:MC_fwf}  
\end{figure}

\begin{figure}[t]
	\centering
	\includegraphics[scale=0.22]{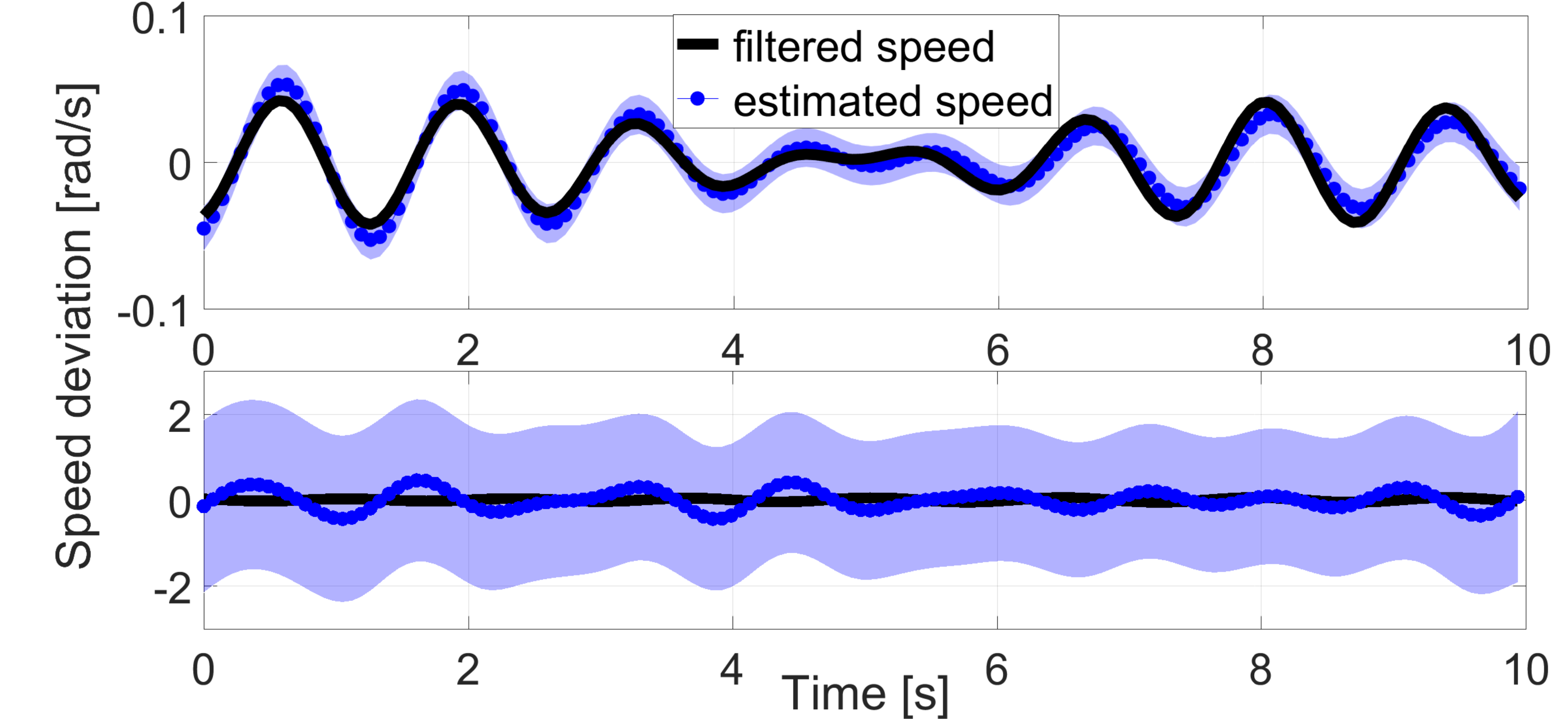}
	\caption{Estimating speeds at buses $31$ (top) and $68$ (bottom) of the $69$-bus system under ambient conditions using speed measurements at buses $\{22-30,32-61\}$. The speed at bus $68$ depends heavily on eigenstate $\dot{y}_4(t)$. This eigenstate depends weakly on the measured speeds. As expected, the estimate of $\omega_{68}(t)$ is unreliable, yet this is signified by the wider uncertainty interval inferred.}	
	\label{fig:obs}  
\end{figure}

The previous experiment was for a single placement of metered buses. To explore how different placements of metered buses affect the estimation accuracy, we performed two additional tests: one with Monte Carlo runs on random placements and another test with an intentionally poor placement. For the first test, we estimated the speeds at non-metered buses using angle or speed measurements collected at $40$, $50$, and $60$ randomly sampled buses. The test was repeated for $100$ different random placements of metered buses. The obtained results are shown in Fig.~\ref{fig:MC_fwf}. The absolute error on estimating speeds was averaged over the non-metered buses and over the Monte Carlo runs of random placements. Fig.~\ref{fig:MC_fwf} demonstrates that on the average, the GP-based estimator attains speed errors smaller than $3.5\cdot 10^{-3}$~rad/sec or $5.5\cdot 10^{-4}$~Hz. As expected, the estimation error decreases with increasing number of metered buses. Of course, plotting results averaged over buses and placements can be hiding severe observability issues. To showcase such issues, we conducted a second test. We metered speeds at buses $\{22-30,32-61\}$. These buses seemed to have little effect on eigenstate $\dot{y}_4(t)$, that is the coefficients associated with these buses on the $4$-th row of matrix $\bV^\top\bM^{1/2}$ in \eqref{eq:trans} were negligible. On the other hand, we identified that speed $\tilde{\omega}_{68}(t)$ depends heavily on $\dot{y}_4(t)$, and $\tilde{\omega}_{31}(t)$ does not. Fig.~\ref{fig:obs} shows the speed estimates for buses $68$ and $31$. The estimate for bus 31 is reliable, whereas the one for bus 68 is not. Interestingly enough though, the proposed method returned very broad confidence intervals for $\hat{\omega}_{68}(t)$, which essentially flag this estimate as unreliable.

\begin{figure}[t]
	\centering
	\includegraphics[scale=0.23]{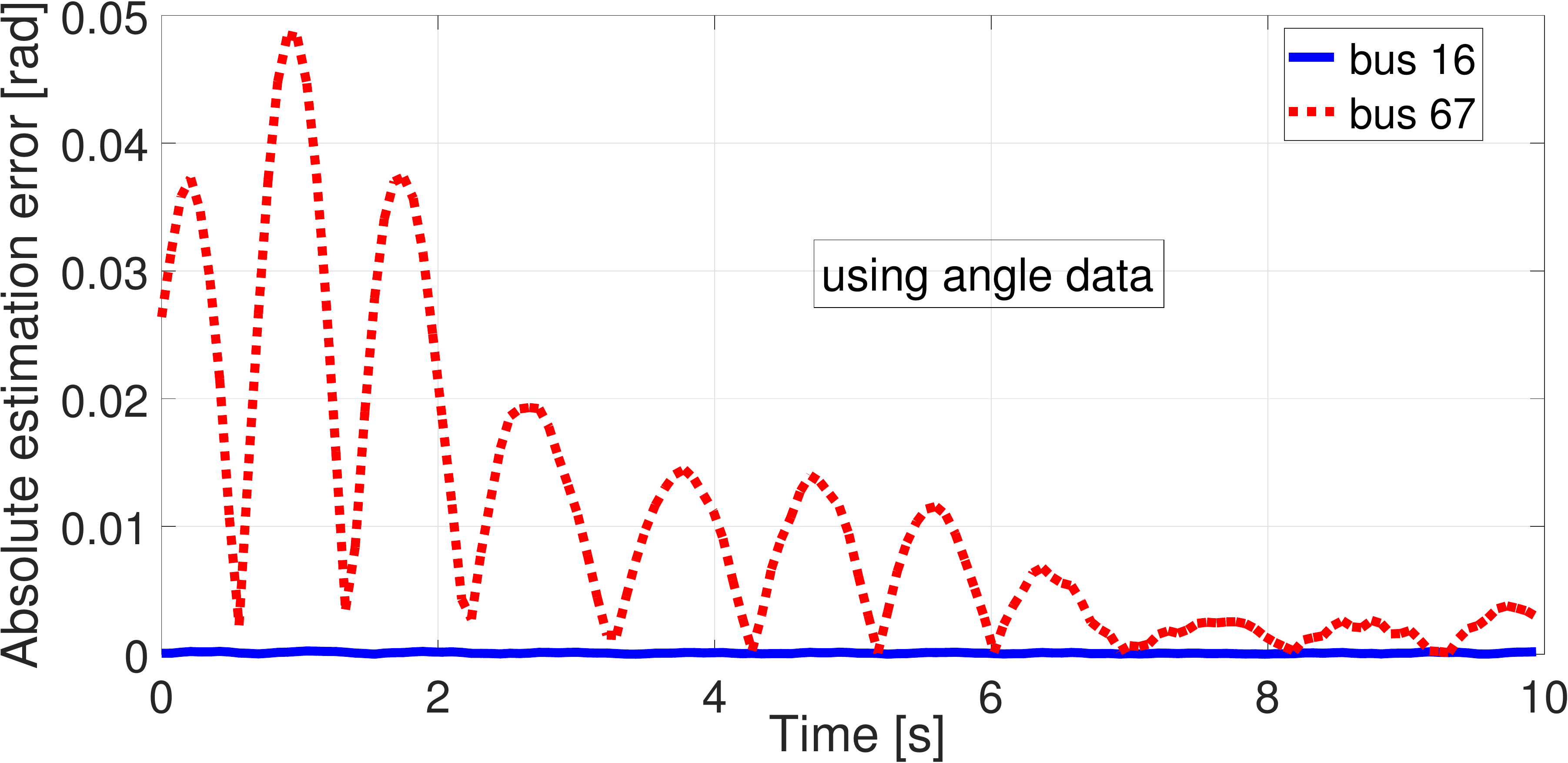}
	\includegraphics[scale=0.24]{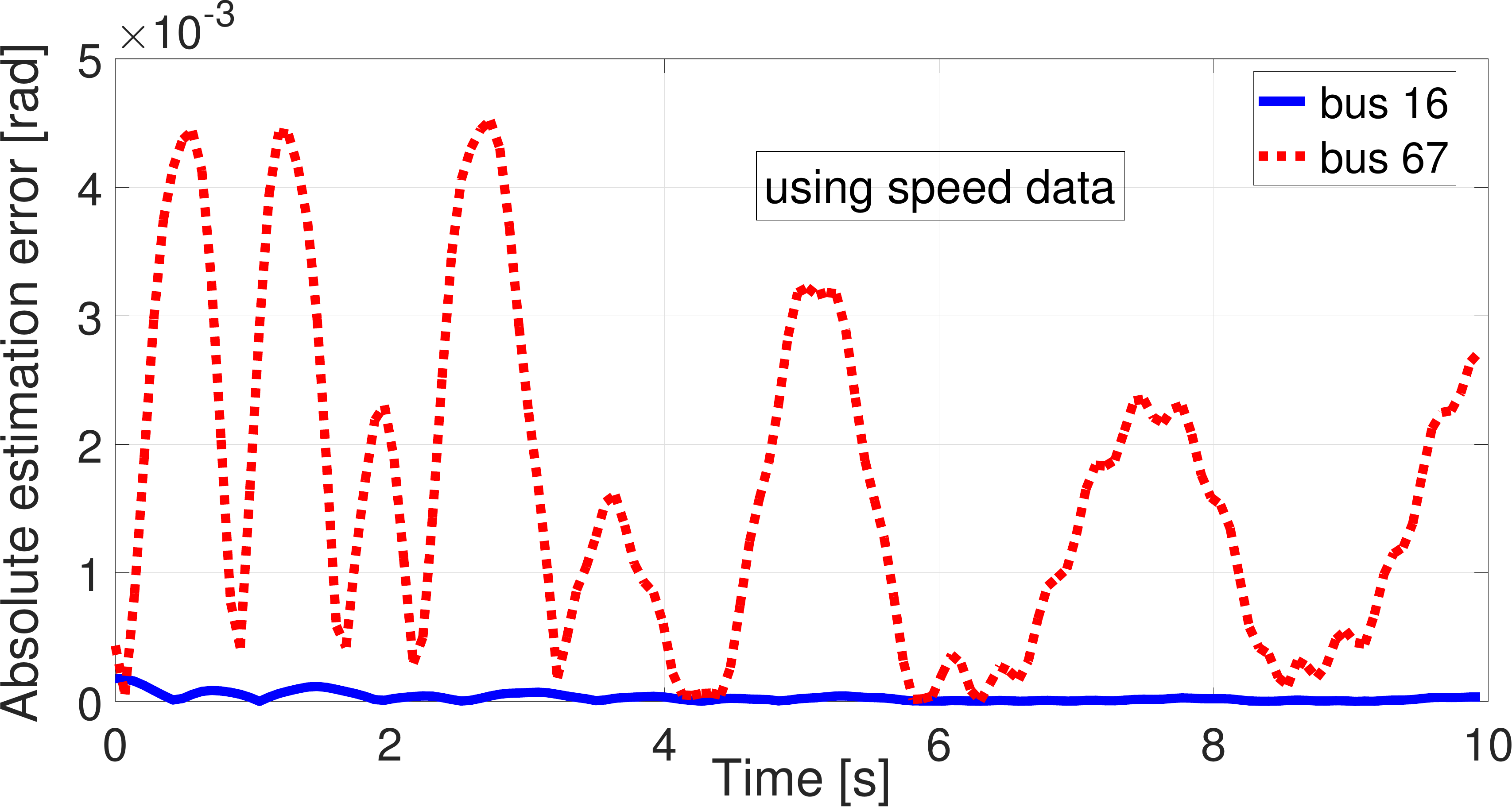}
	\caption{{Estimating angle dynamics following a generator trip using angle (top) and speed (bottom) data metered at 50 buses. The plot shows absolute estimation errors for the buses with the smallest/largest mean absolute error (MAE) of $E_n$.}}
	\label{fig:error_impulse_twt}  
\end{figure}

\begin{figure}[t]
	\centering
	\includegraphics[scale=0.21]{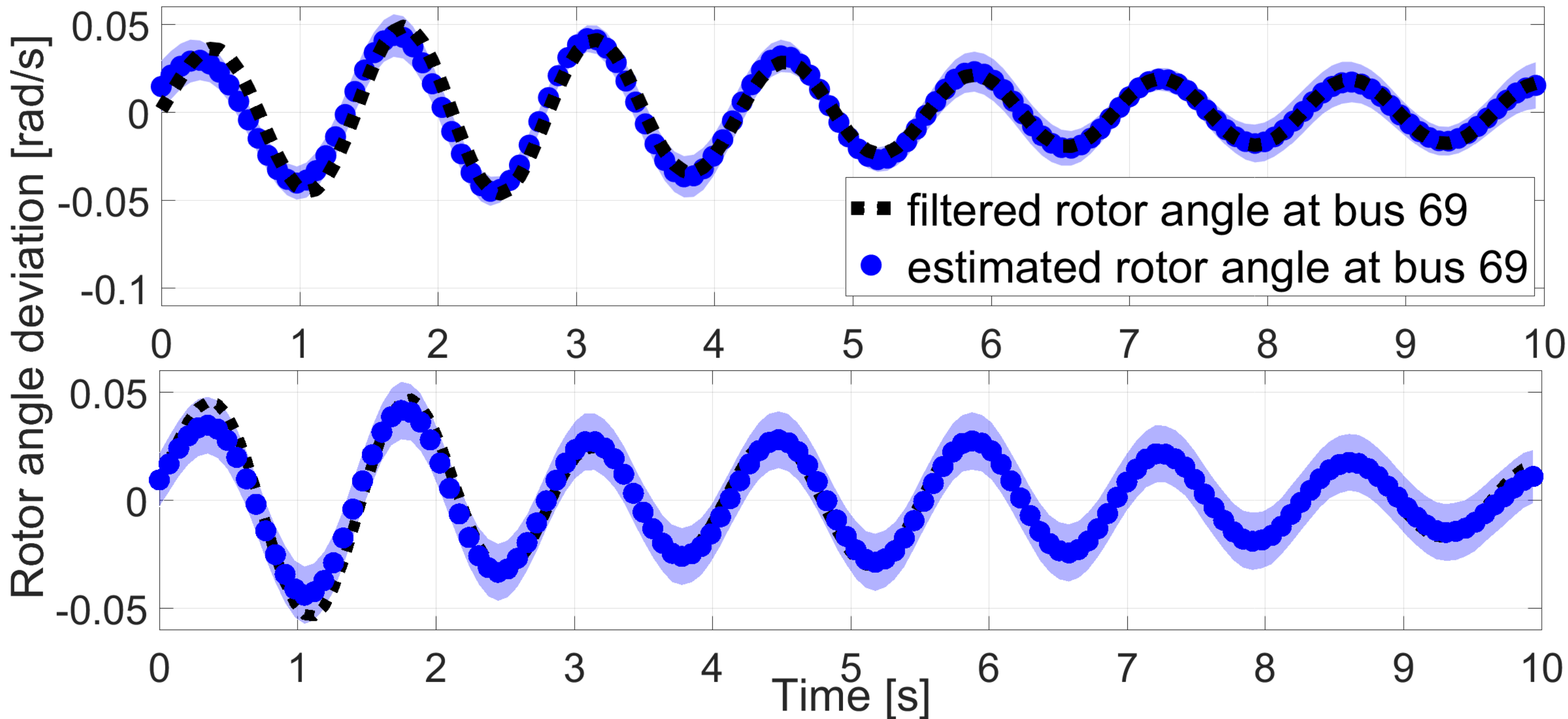}
	\caption{Estimating the angle at bus 69 of the 69-machine system after a fault near the same bus using angle (top) and speed (bottom) measurements from $50$ buses.}	
	\label{fig:L_theta}  
\end{figure}

Next, we aimed at finding angles under a fault at generator $69$. We used angle and frequency data at $50$ buses with bus $69$ apparently excluded. Fig.~\ref{fig:error_impulse_twt} depicts the absolute error in angle estimates for the buses with the largest {(bus $67$)} and smallest {(bus $16$)} time-averaged errors. Fig.~\ref{fig:L_theta} showcases that GP learning can successfully infer angles from speeds, and can deal with different types of disturbances. The latter conclusion is important because the correlation among system states depends on the type and magnitude of disturbances. Thanks to the parametric covariance model using $\tilde{\bA}$, the GP framework is adaptable to different dynamic conditions. 

\begin{figure}[t]
	\centering
	\includegraphics[scale=0.21]{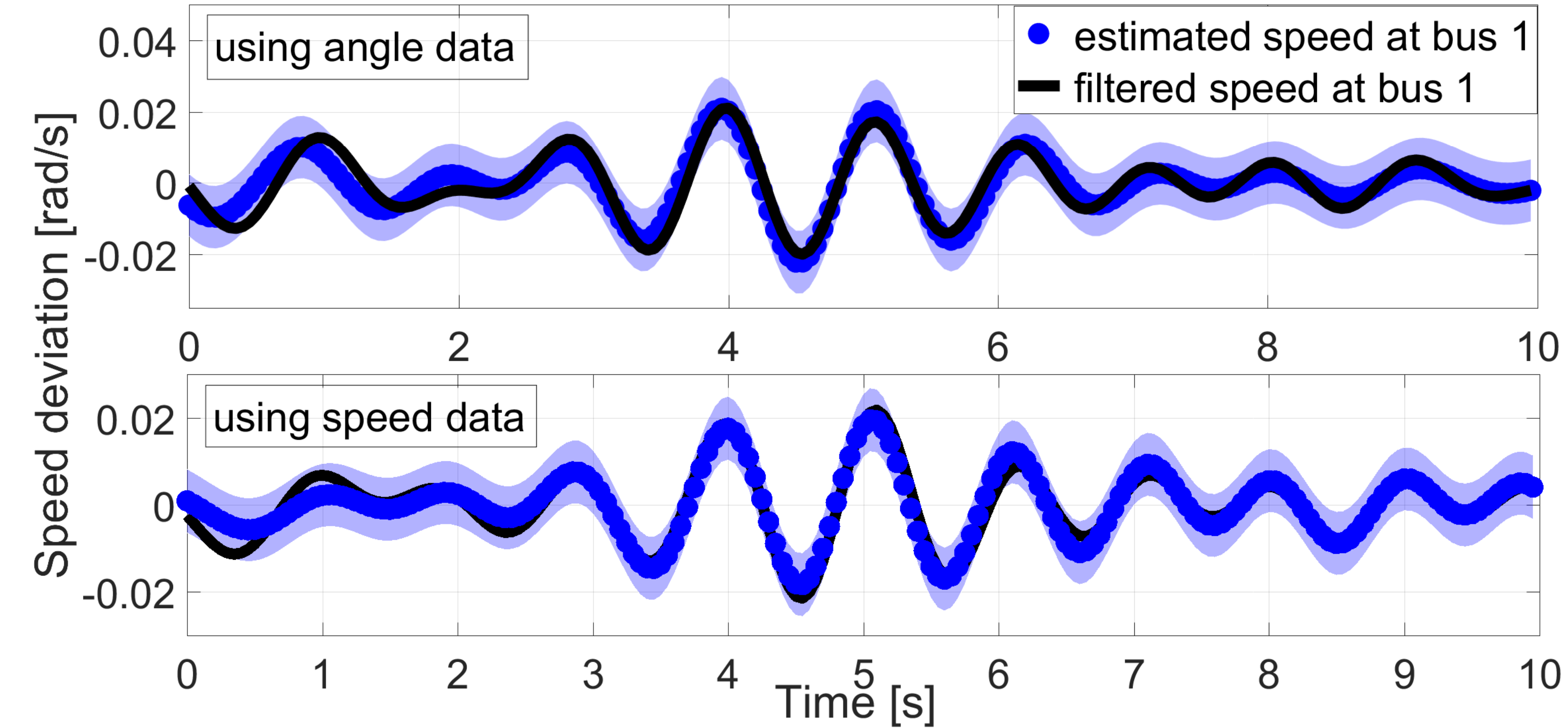}
	\caption{Speed estimates at bus $1$ of the nonlinear higher-order system model \emph{M2)} under ambient dynamics using angle (top) and speed (bottom) data of the system except buses $\{1,5,7\}$}
	\label{fig:N_freq}  
\end{figure}

\subsubsection{Inferring Rotor Angles and Speeds Using M2)}
\begin{figure}[tp]
	\centering
	\includegraphics[scale=0.21]{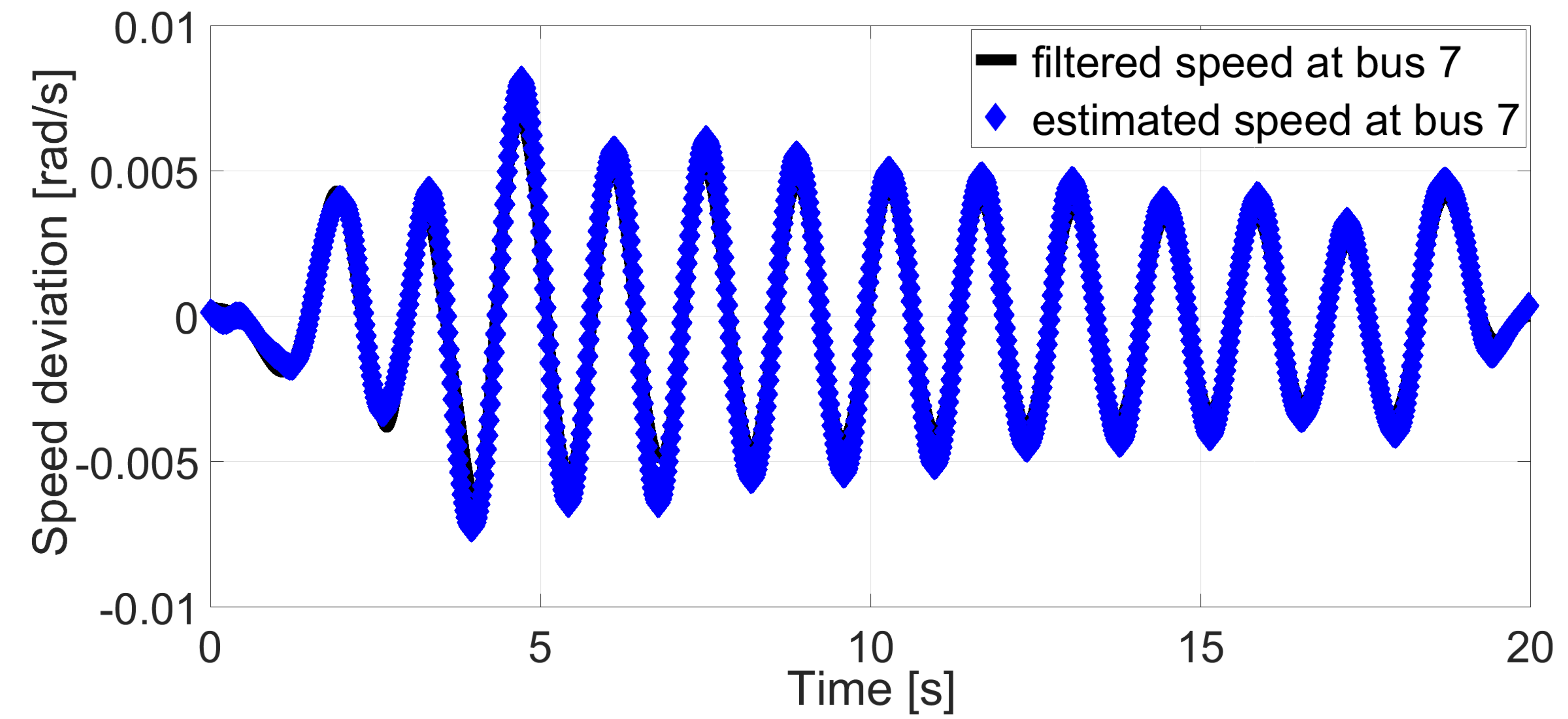}
	\includegraphics[scale=0.21]{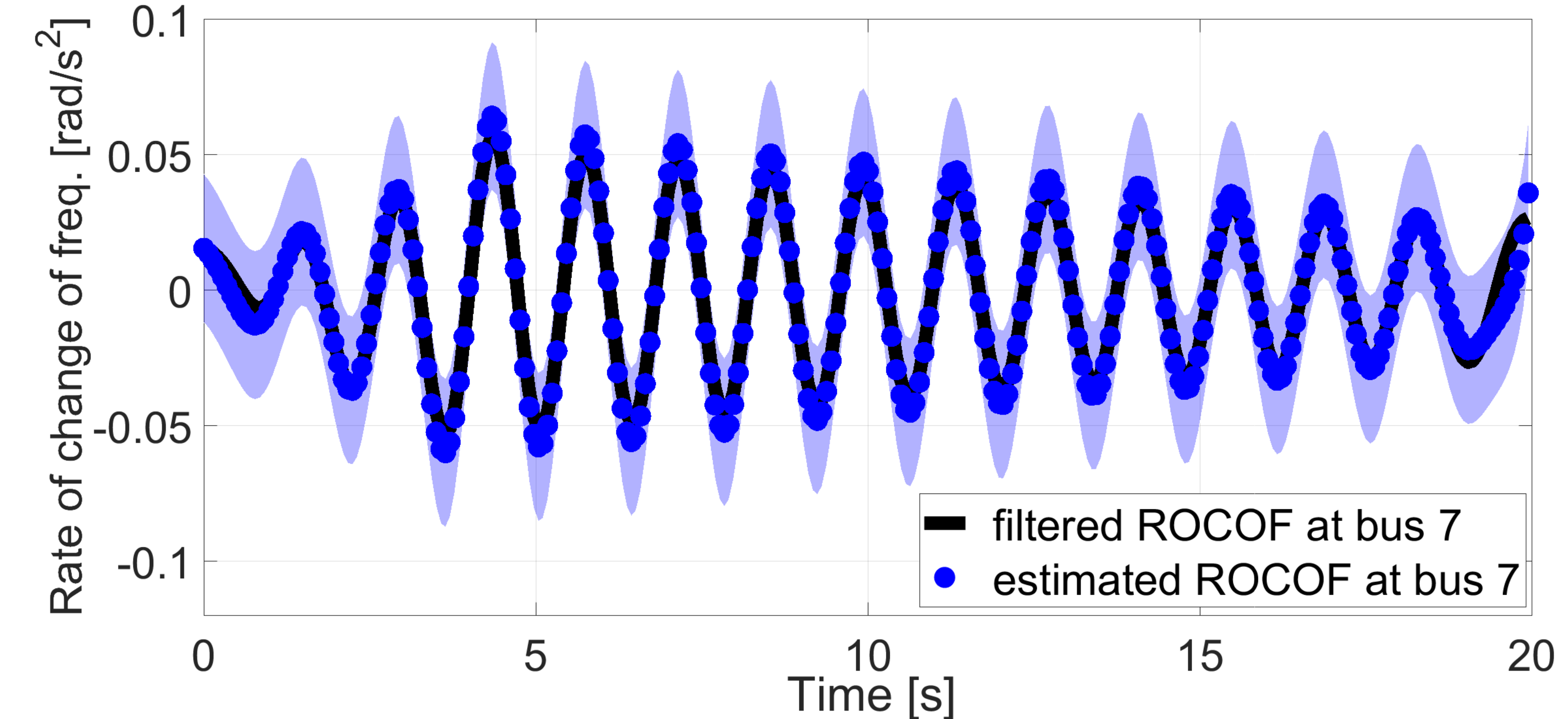}
	\caption{Speed (top) and ROCOF (bottom) estimates at bus $7$ of the nonlinear system model \emph{M2)} under a $3$-phase fault near bus $7$ using speed measurements across the system except buses $\{3,6,7\}$.}	
	\label{fig:N_fwf_fault}  
\end{figure}

We next evaluated GP learning using data generated by the nonlinear higher-order model~\emph{M2)} under ambient and non-ambient conditions. Under ambient conditions, angle and speed data were collected at all buses except $\{1,5,7\}$. For non-ambient conditions, we simulated a $3$-phase fault near bus $7$, and collected angle speed at all buses except $\{3,6,7\}$. Figs.~\ref{fig:N_freq} and Fig.~\ref{fig:N_fwf_fault} illustrate the inferred oscillations in rotor speeds and ROCOFs at non-metered buses. The results in Fig.~\ref{fig:N_fwf_fault} corroborate that although the parametric form of covariances was designed under the stylized model of Section~\ref{sec:power} (linearized swing dynamics with uniform damping excited by specific Gaussian perturbations), GP learning yields accurate results when trained and operated over data from a higher-order nonlinear model with non-uniform damping excited by (non)-ambient perturbations. 

\subsubsection{Inferring Rotor Speeds using M3)} 
\begin{figure}[t]
	\centering
	\includegraphics[scale=0.23]{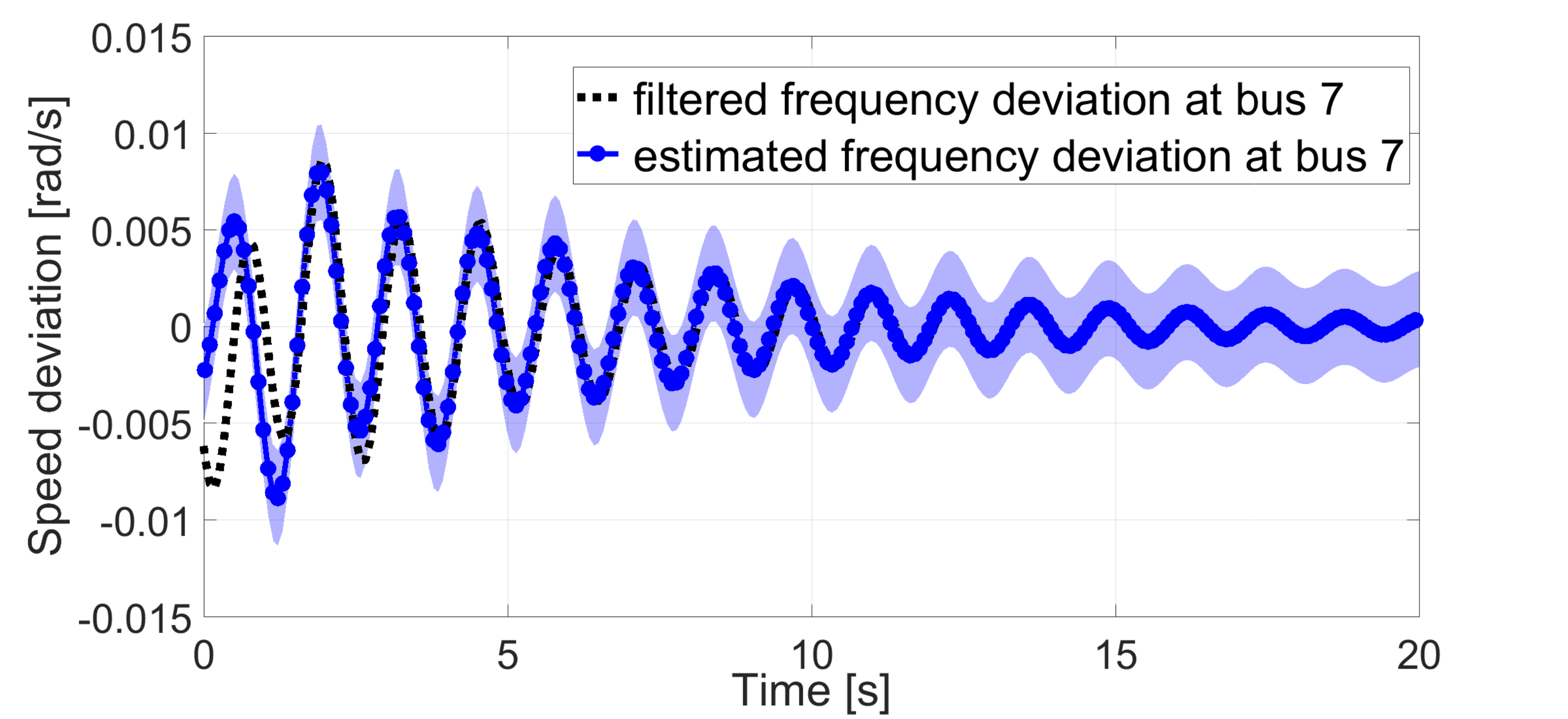}
	\caption{{Speed estimation at bus $7$ of the $10$-bus system under a fault near bus $2$ using speed measurements.}}	
	\label{fig:f7_TG}  
\end{figure}

We also estimated speed oscillations using model \emph{M3)}, which includes turbine/droop control. We simulated a fault near generator $2$, and measured all buses except for $\{6,7\}$. Non-metered speeds were found using the covariance model of Proposition~\ref{prop:turbine}. Fig.~\ref{fig:f7_TG} shows the estimation results at bus $7$. This result proves that the physics-informed covariance model can be modified to capture other dynamic components of the system.

\begin{figure}[t]
	\centering
	\includegraphics[scale=0.22]{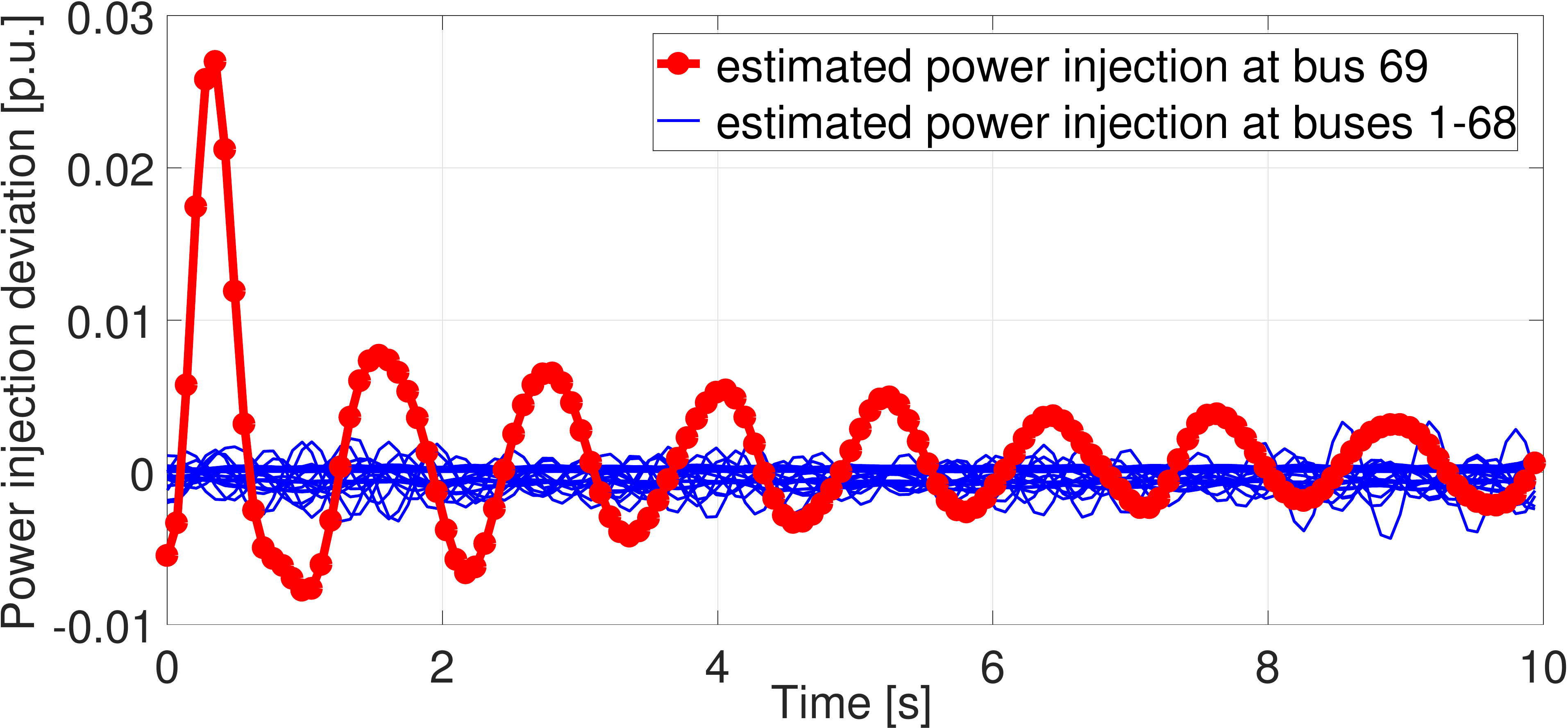}
	\caption{GP learning can be used to infer power injections given angle or speed measurements. Here we estimate the power injection at each bus upon processing speed data at all buses following a fault near bus $69$.}	
	\label{fig:p}  
\end{figure}

\subsection{Localizing Sources of Oscillations using M1):}\label{subsection:ii}
So far, we have focused on inter-area oscillations. However, in some applications such as fault identification, the eigenstate corresponding to the zero eigenvalue is required. This is because this eigenstate is directly related to the system frequency, also known as \emph{center of inertia}. To test the ability of GP learning to locate faults, we tried estimating the power injections across the $69$-bus system under a fault near generator $69$. To this end, we passed speed measurements collected at 50 buses through a low-pass filter with a cut-off frequency of $2$~Hz ($\Omega=[0,2]$~Hz). Fig.~\ref{fig:p} illustrates the estimated power injections at all buses. The results show that the GP framework can successfully localize the bus where the fault occurred. Since we are working with filtered data, the recovered power injections shown in Fig.~\ref{fig:p} correspond to the filtered power injections, and hence, the impulse-like fault injection at bus $69$ appears as a smoothed sync-type form and of reduced magnitude; the nominal value of the active power injection at the faulted bus was $p_{69}=0.08$~pu. Nonetheless, it is still very clear that the disturbance occurred at bus $69$.

\begin{figure}[t]
	\centering
	\includegraphics[scale=0.22]{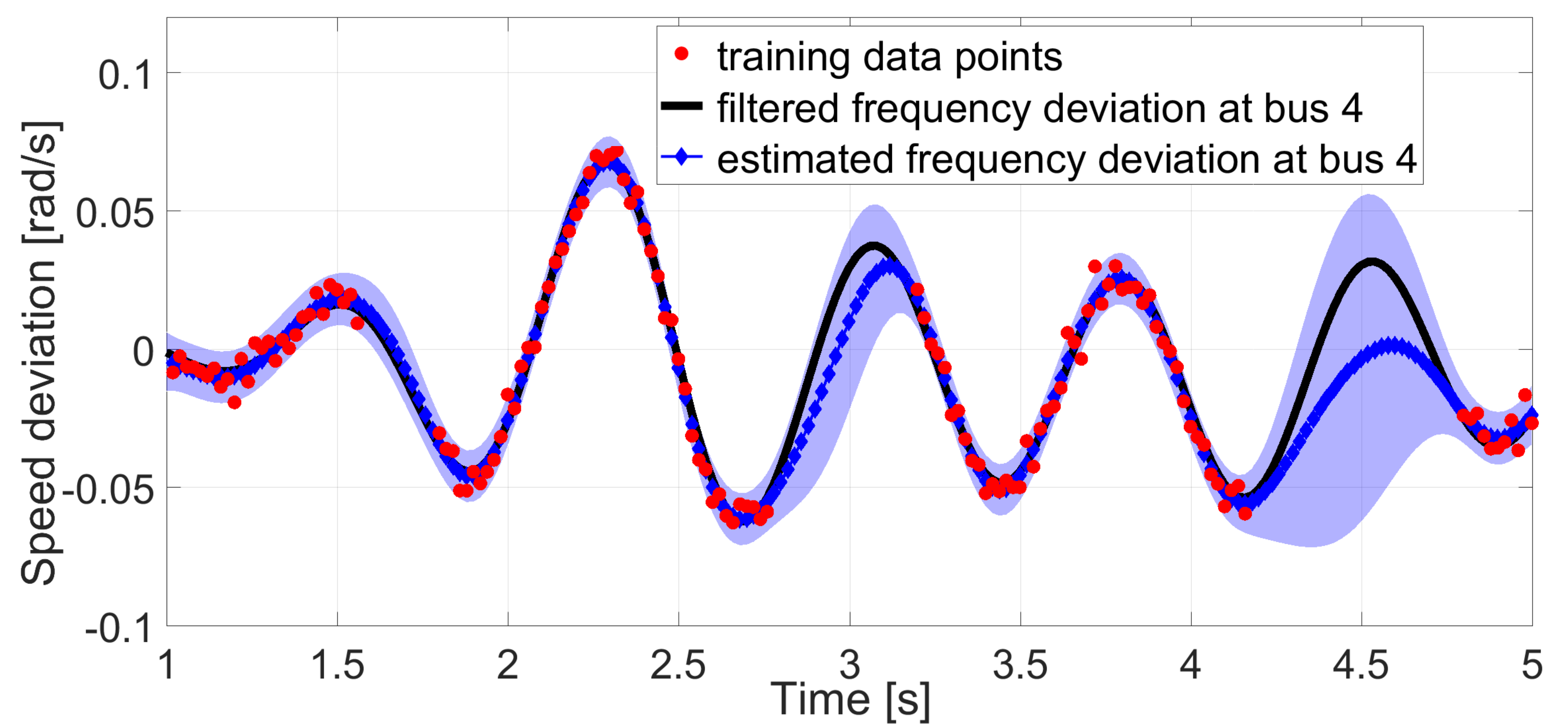}
	\caption{Model-free imputation of missing data for the speed of generator $4$ of the $10$-bus system. For periods of missing data of shorter duration, estimation accuracy improves and uncertainty decreases.}	
	\label{fig:interp}  
\end{figure}

\subsection{Model-free Imputation of Missing Data across Time}\label{subsec:iii}
The previous experiments focused on learning oscillations at non-metered buses. To be able to extrapolate across buses, we used the covariance model of~\eqref{eq:cov3}, which relies on an approximate dynamic model of the power system. Nevertheless, in applications where one processes a single grid signal, there is no need of knowing the power system model; temporal covariances of the form $\mathbb{E}[\omega_n(t) \omega_n(t+\tau)]$ suffice. To see one such application, consider a dynamic speed signal from bus $4$ of model~\emph{M2)} under a three-phase fault near generator $2$. Due to communication failures, some samples over stretches of consecutive time instances are missing. We aim at inferring the missing data using the available speed data from bus $4$. Focusing on inter-area oscillations, we filter the measurements using $\Omega=[0.5,0.8]$~Hz. In this setup, we learn a Gaussian (temporal only) covariance for rotor speeds. Its parameters used in the Gaussian kernel were estimated using MATLAB's \texttt{fitRGP} toolbox. Fig.~\ref{fig:interp} demonstrates the imputed oscillations over three intervals of increasing duration. As expected, estimation improves and uncertainty decreases when the missed interval is shorter.

\begin{figure}[t]
	\centering
	\includegraphics[scale=0.22]{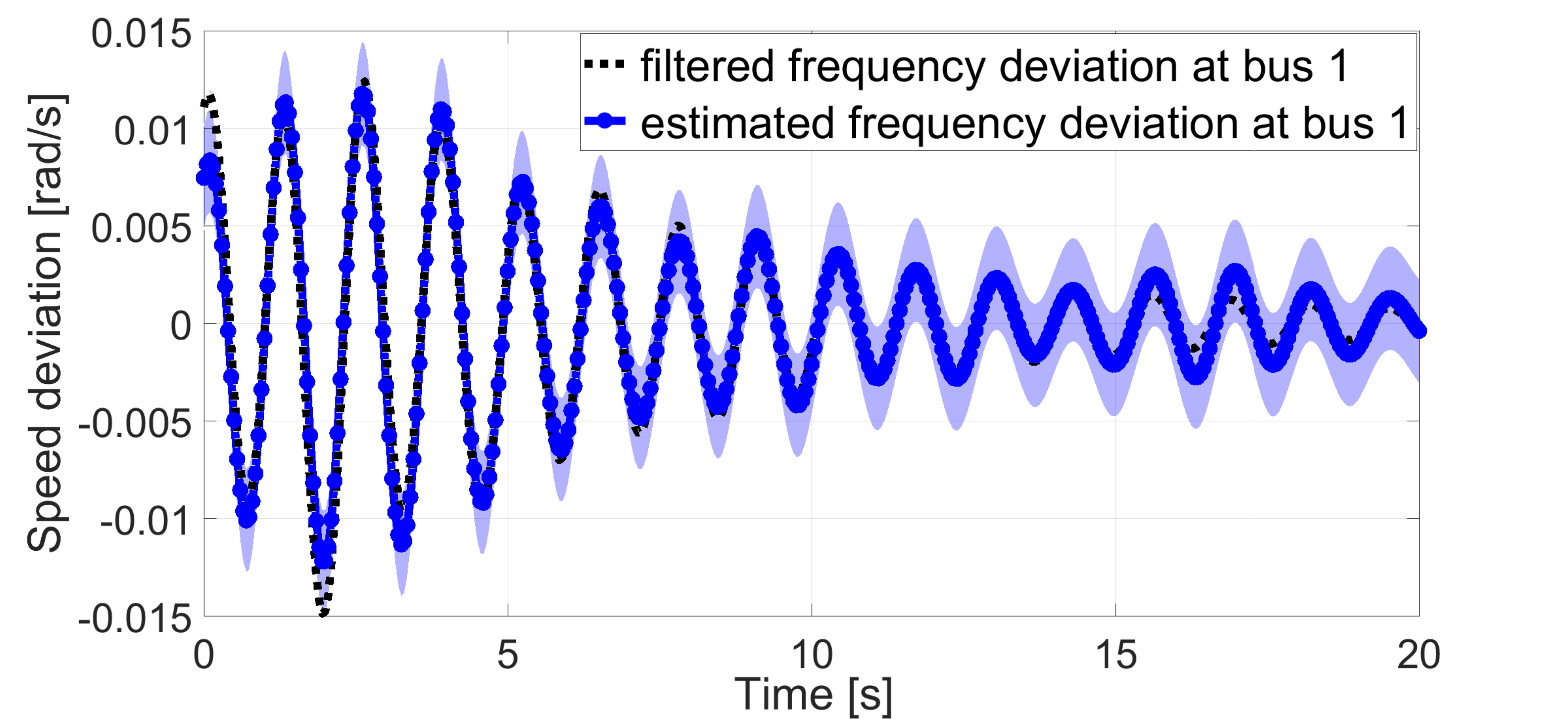}
	\caption{{Model-free rotor speed estimation at bus $1$ of the $10$-bus system.}}	
	\label{fig:model_free_fwt}  
\end{figure}

\subsection{Model-free Imputation of Rotor Speeds from Angle Data}\label{subsec:iv}
Finally, we used GP learning to compute speeds given angle measurements collected at the same bus. In essence, GP learning is used here in lieu of time differentiation. This could serve as an alternative to the Fourier-based frequency estimation techniques implemented by PMU devices. We assumed that rotor angle data are available at generator $1$ of model~\emph{M3)} under a fault near generator $2$. We modeled angle covariances using the Gaussian covariance functions. Rotor covariances can be derived as discussed in~\eqref{eq:kdot}. The rotor speed estimation results are shown in Fig.~\ref{fig:model_free_fwt}. The results confirm that model-free covariance functions can be used with the proposed GP framework for implementing time differentiation. It can be observed that the uncertainty interval lies within $\pm 0.0025$~rad/s or $\pm 0.0004$~Hz. Therefore, $3\sigma=0.0012$~Hz is less than the maximum frequency measurement error specified by the IEEE C37118 Standard for PMU measurements~\cite{C37118}.

\section{Conclusions and Future Work}\label{sec:conc}
A novel method for inferring the non-metered dynamic oscillations in a power system using synchrophasor data has been put forth. The key idea has been to capture voltage frequencies as GPs and systematically propagate this GP model to voltage angles, speeds, ROCOFs, and power injections. Leveraging information on the power network model and generator parameters, the proposed GP framework can interpolate and extrapolate dynamic grid signals across buses and time. It can process synchrophasor data with diverse characteristics, such as sampling rate, type (angles, speeds/frequencies, ROCOFs), and accuracy, or with missing entries. Signals corresponding to time derivatives can be learned by analytically differentiating kernel functions rather than approximating them using finite differences. Due to its Bayesian nature, the proposed model provides confidence intervals in addition to point estimates.

Although the statistical model was developed on linearized dynamics presuming uniform damping, numerical tests on the IEEE 300- and 39-bus benchmarks have corroborated that the method performs well under non-uniform and/or nonlinear system models under both ambient and fault conditions. The tests have shown how: \emph{i)} one can estimate speeds or ROCOF and to locate faults using angle and speed data; \emph{ii)} accuracy improves with increasing number of measurements and remains acceptable in general for a random meter placement; \emph{iii)} observability issues can arise and are identified by the uncertainty estimates provided by the method. 

This work sets the foundations for interesting research directions. The online implementation of the method, frequency prediction, system model estimation, and modal analysis are a few practically pertinent extensions. Finally, exploring more detailed generator models and using other measurements such as field voltages and line flows could improve estimation accuracy. 
\bibliographystyle{IEEEtran}
\bibliography{myabrv,power}
\end{document}